\documentclass[twocolumn]{aastex63}

\newcommand\aastex{AAS\TeX}

\usepackage[figuresright]{rotating}
\usepackage{amsmath}
\usepackage{amsfonts}
\usepackage{graphicx}
\usepackage{rotating}
\usepackage{natbib}
\usepackage{txfonts}
\usepackage{color}
\usepackage{wrapfig}
\usepackage{amssymb}
\usepackage{color}
\usepackage{xcolor}
\usepackage{ulem}

\received{2017 July 10}
\revised{2017 October 12}
\accepted{2017 October 27}

\submitjournal{ApJ}

\shorttitle{\aastex\ Interstellar aldimines and amines}
\shortauthors{Sil et al.}

\begin{document}

\title{Chemical modeling for predicting the abundances of certain aldimines and amines in hot cores}
\correspondingauthor{Ankan Das}
\email{ankan.das@gmail.com}
 
\author[0000-0001-5720-6294]{Milan Sil}
\affiliation{Indian Centre for Space Physics, 43, Chalantika, Garia Station Road, Kolkata 700084, India}
\author[0000-0003-1602-6849]{Prasanta Gorai}
\affiliation{Indian Centre for Space Physics, 43, Chalantika, Garia Station Road, Kolkata 700084, India}
\author[0000-0003-4615-602X]{Ankan Das}
\affiliation{Indian Centre for Space Physics, 43, Chalantika, Garia Station Road, Kolkata 700084, India}
\author[0000-0002-5224-3026]{Bratati Bhat}
\affiliation{Indian Centre for Space Physics, 43, Chalantika, Garia Station Road, Kolkata 700084, India}
\author[0000-0001-8304-9771]{Emmanuel E. Etim}
\affiliation{Department of Chemical Sciences, Federal University Wukari, Katsina-Ala Road, P.M.B. 1020 Wukari, Taraba State, Nigeria}
\author[0000-0002-0193-1136]{Sandip K. Chakrabarti}
\affiliation{S. N. Bose National Centre for Basic Sciences, Salt Lake, Kolkata 700098, India}
\affiliation{Indian Centre for Space Physics, 43, Chalantika, Garia Station Road, Kolkata 700084, India}

\begin{abstract}
We consider six isomeric groups ($\rm{CH_3N,~CH_5N,~C_2H_5N,~C_2H_7N, ~C_3H_7N}$ and $\rm{C_3H_9N}$)
to review the presence of  amines and aldimines within the interstellar medium (ISM).  
Each of these groups contains
at least one aldimine or amine. Methanimine ($\mathrm{CH_2NH}$) from $\rm {CH_3N}$ and 
methylamine ($\mathrm{CH_3NH_2}$) from $\rm{CH_5N}$ isomeric group were detected a few decades ago. 
Recently, the presence of 
ethanimine ($\mathrm{CH_3CHNH}$) from $\rm{C_2H_5N}$ isomeric group has been discovered in the ISM. 
This prompted us to investigate the possibility of detecting 
any aldimine or amine from the very next three isomeric groups in this sequence: $\rm{C_2H_7N}$,
$\rm{C_3H_7N}$ and $\rm{C_3H_9N}$. We employ high-level quantum chemical calculations 
to estimate accurate energies of all the species.  
According to enthalpies of formation, optimized energies,
and expected intensity ratio, 
we found that ethylamine (precursor of glycine)
from $\mathrm{C_2H_7N}$ isomeric group, (1Z)-1-propanimine  
from $\rm{C_3H_7N}$ isomeric group, and trimethylamine from $\rm{C_3H_9N}$ isomeric group 
are the most viable candidates for the future astronomical detection. 
Based on our quantum chemical calculations and from other approximations 
(from prevailing similar types of reactions), a complete set of reaction pathways to the 
synthesis of ethylamine and 
(1Z)-1-propanimine is prepared. Moreover, a large gas-grain chemical model is employed to 
study the presence of these species in the ISM. Our modeling results suggest that ethylamine 
and (1Z)-1-propanimine could efficiently be formed in hot-core regions and could be observed with present
astronomical facilities. 
Radiative transfer modeling is also implemented to additionally aid their discovery 
in interstellar space.
\end{abstract}

\keywords{astrochemistry -- dust, extinction -- evolution -- ISM: abundances -- ISM: molecules -- methods: numerical \\
\\
\\
\textit{Supproting materials:} tar.gz files}

\section{Introduction}
Almost $200$ species have been discovered in the interstellar medium (ISM) or 
circumstellar shells and most of them are organic in nature (\url{https://www.astro.uni-koeln.de/cdms/molecules}). The presence of numerous complex 
organic molecules in the ISM has been reported earlier \citep{herb06,cron93}.
The chemical richness of extraterrestrial media points to the formation of biomolecules in the ISM. 
There are some studies on the formation of complex pre-biotic molecules in collapsing clouds
and star-forming regions \citep{garr13,chak00a,chak00b,chak15}. It is suggested 
that if the abundance of bio-molecules is too small for detection, their precursors may be
observed for estimating the abundances of these bio-molecules \citep{maju12,maju13}.
Aldimines and amines are the building blocks of amino acids \citep{godf73,holt05}. 
Thus the discovery of these species under astrophysical
circumstances could be treated as important clues leading to the origin of life.

Aldimines are very important as they are seen within the reactions of Strecker-type synthesis. 
The Strecker synthesis prepares $\alpha$-aminonitriles, which are versatile 
intermediates for the synthesis of amino acids via hydrolysis of nitriles. However, a Strecker-type 
formation route was found to be less important \citep{elsi07}. 
Therefore, we investigate a total of $34$ molecules from six isomeric groups 
namely, $\mathrm{CH_3N}$, ~ $\mathrm{CH_5N}$, 
~$\mathrm{C_2H_5N}$, ~$\mathrm{C_2H_7N}$, $\rm{C_3H_7N}$ and $\mathrm{C_3H_9N}$
to find out the possibility of detecting some of the aldimines and amines which are the
precursor of amino acids. From these six isomeric 
groups, at least one species was observed from each of the $\rm{CH_3N}$, $\rm{CH_5N}$ and $\rm{C_2H_5N}$ 
isomeric groups. However, to date, any species from the $\rm{C_2H_7N}$, $\rm{C_3H_7N}$
and $\rm{C_3H_9N}$ isomeric groups is yet to be detected.

From the $\mathrm{CH_3N}$ isomeric group, methanimine ($\mathrm{CH_2NH}$) 
was observed in Sgr B2 \citep{godf73}.
The simplest amino acid, namely, glycine could have been formed by 
the reaction between methanimine and formic acid \citep{godf73}. 
In that sense, methanimine is the precursor molecule for glycine \citep{suzu16}. 
Similarly, from the $\mathrm{CH_5N}$ isomeric group, methylamine ($\mathrm{CH_3NH_2}$) was detected 
\citep{kaif74, four74} in both Sgr B2 and Orion A. 
\cite{holt05} showed from theoretical and experimental studies that glycine
could have been formed by the reaction with another precursor molecule (reaction between 
methylamine and $\mathrm{CO_2}$ under UV irradiation on an icy grain mantle). On the other hand, methylamine 
could be produced by two successive hydrogen addition reactions with methanimine.
It can also be formed by four successive hydrogen additions to HCN on the 
surface of grains \citep{godf73, woon02, theu11}.
Both the precursor molecules (methylamine and ethylamine) of glycine were 
observed in comet wild 2 \citep{glav08} and in the coma of 67P/Churyumov-Gerasimenko 
(by the Rosetta Orbiter Spectrometer for Ion and Neutral Analysis mass spectrometer \citep{altw16}).

Microwave and mm-wave spectra of the two conformers of ethanimine (E- and Z-ethanimine) 
were characterized in order to guide the astronomical searches \citep{lova80, brow80}. 
They recommended that ethanimine from the $\rm{C_2H_5N}$ isomeric group  
should be a possible interstellar molecule which can be seen in space. 
Finally, ethanimine has been detected with both the forms 
in the same sources where methanimine has
already been observed \citep{loom13}.

From the $\rm{C_2H_7N}$, $\rm{C_3H_7N}$ and $\rm{C_3H_9N}$ isomeric groups, 
ethylamine, propanimine and trimethylamine are of special interests because 
they could possibly play a role towards the formation of amino acids and other pre-biotic molecules. 
Recently, \cite{marg15} has performed the first 
spectroscopic study of propanimine molecule. Similar to the ethanimine molecule, 
\cite{marg15} found that propanimine can exist in two conformations, 
E-propanimine and Z-propanimine. In order to detect these species under
astrophysical conditions, the spectroscopic details as well as the chemical abundances 
of these species are essential to know.

Section 2 describes the computational details and methodology of 
our work. Results are extensively discussed in Section 3. General conclusions of this
work are provided in Section 4. In the Appendix, we present
the gas phase formation and destruction pathways. 
As the supplementary materials of this work, we provide the
catalog files for ethylamine and (1Z)-1-propanimine which could be
very useful for the future detection of these species in the ISM.

\section{Computational Details and Methodology}
\subsection{Quantum chemical calculations}
All the calculations are performed with the Gaussian 09 suite of programs \citep{fris13}. 
Table 1 shows some ice phase reactions which lead to the formation of 
various interstellar amines and aldimines. Some of these reactions are
radical-radical (RR) in nature and thus can happen at each encounter. However, there are
some reactions between neutrals and radicals (NR) which often possess 
activation barriers. The QST2 method with 
B3LYP/6-311++G(d,p) 
levels of theory are employed
to calculate various energy barriers (activation barrier and Gibbs free energy of activation).
Moreover, QST2 method is also used to determine reaction pathways and transition state structures.

\begin{deluxetable*}{ccc}
\tablecaption{Ice phase formation pathways.
\label{table:reactions}}
\tablewidth{0pt}
\tabletypesize{\scriptsize}
\tablehead{
\colhead{\bf Reaction number (type)} & \colhead{\bf Reactions} & \colhead{\bf Activation barrier (K)}}
\startdata
R1(RR)$^a$&$\mathrm{N + CH_3 \rightarrow CH_2NH}$&0.0\\
R2(RR)$^a$&$\mathrm{NH  + CH_2 \rightarrow CH_2NH}$&0.0\\
R3(RR)$^a$&$\mathrm{NH_2  + CH \rightarrow CH_2NH}$&0.0\\
R4(NR)$^a$&$\mathrm{HCN+H \rightarrow H_2CN}$&3647$^e$ \\
R5(NR)$^a$&$\mathrm{HCN+H \rightarrow HCNH}$&6440$^e$ \\
R6(RR)$^a$&$\mathrm{H_2CN+H \rightarrow CH_2NH}$&0.0\\
R7(RR)$^a$&$\mathrm{HCNH+H \rightarrow CH_2NH}$&0.0\\
R8(NR)$^a$&$\mathrm{CH_2NH + H \rightarrow CH_3NH} $&2134$^e$ \\
R9(NR)$^a$&$\mathrm{CH_2NH + H \rightarrow CH_2NH_2}$&3170$^e$ \\
R10(RR)$^a$&$\mathrm{CH_3NH+H \rightarrow CH_3NH_2}$&0.0\\
R11(RR)$^a$&$\mathrm{CH_2NH_2+H \rightarrow CH_3NH_2}$&0.0\\
R12(RR)$^b$&$\mathrm{CH_2CN+H \rightarrow CH_3CN}$&0.0\\
R13(RR)$^c$&$\mathrm{CH_3+CN \rightarrow CH_3CN}$&0.0\\
R14(NR)$^d$&$\mathrm{CH_3CN+H \rightarrow CH_3CNH}$& 1400$^e$ \\
R15(RR)$^d$&$\mathrm{CH_3CNH+H \rightarrow CH_3CHNH}$&0.0\\
R16(RR)$^d$&$\mathrm{CH_3+H_2CN \rightarrow CH_3CHNH}$&0.0\\
R17(NR)&$\mathrm{CH_3CHNH+H \rightarrow CH_3CH_2NH}$& 1846\\
R18(RR)&$\mathrm{CH_3CH_2NH+H \rightarrow CH_3CH_2NH_2}$&0.0\\
R19(RR)&$\mathrm{C_2H_5+H_2CN \rightarrow CH_3CH_2CHNH}$&0.0\\
R20(RR)&$\mathrm{C_2H_5+CN \rightarrow CH_3CH_2CN}$&0.0\\
R21(NR)&$\mathrm{CH_3CH_2CN+H \rightarrow CH_3CH_2CNH}$& 2712 \\
R22(RR)&$\mathrm{CH_3CH_2CNH+H \rightarrow CH_3CH_2CHNH}$&0.0\\
\enddata
\tablecomments{$^a$ \cite{suzu16}, $^b$ \cite{hase92}, $^c$ \cite{quan10}, $^d$ \cite{quan16}, $^e$ \cite{woon02}.}
\end{deluxetable*}

In order to estimate accurate enthalpies of formation of all the species of various 
isomeric groups, the Gaussian G4 composite method is used. In arriving at an accurate total
energy for a given species, the G4 composite method performs a sequence of well-defined 
ab-initio molecular calculations \citep{curt07,etim16}. 
Each fully optimized structure is 
verified to be a stationary point (having non-negative frequency) by harmonic 
vibrational frequency calculations. For computing the enthalpy of formation, 
we calculate atomization energy of molecules. Experimental values of the 
enthalpy of formation of atoms are taken from \cite{curt97}. 
In Table 2, we summarize the present astronomical status and enthalpy of 
formation ($\Delta_fH^O$) of all the species considered here. Subsequently, in all the tables, 
we arrange the species according to the ascending order of 
the enthalpy of formation. 
Some experimental values of the enthalpy of formation (if available) are also shown 
for comparison. Relative energies of each isomeric group members
are also shown with G4 level of theory. \cite{osmo07} found that this level of theory is also suitable for the computation
of the enthalpies of formation. Moreover, in Table 2, we also include our calculated 
enthalpies of formation with the B3LYP/6-31G(d,p) level of theory. 
In our case, we found that the calculated enthalpies of 
formation with B3LYP/6-31G(d,p) level of theory are closer to the experimentally obtained 
values than that of the G4 composite method.

Species with permanent electric dipole moments are generally detected from their rotational transitions and
about 80\% of all the known interstellar and circumstellar molecules were discovered by these
transitions. Intensity of any rotational transition is especially dependent on the temperature and
the components (a-type, b-type and c-type) of dipole moment \citep{mcml14,fort14}.
Relative signs of the dipole moment components may cause the 
change of intensities of some transitions \citep{mull16}. These intensities are directly 
proportional to the square of the dipole moment and inversely proportional to the rotational partition
function. Thus, in general, for a fixed temperature, higher the dipole moments, intensities are
higher.
All the molecules considered here have a non-zero permanent electric 
dipole moment. Dipole moment components along the
inertial axis ($\mu_a,\mu_b$ and $\mu_c$) are summarized in Table 3.
We use input and principal axis orientation geometry defined according to the Pickett program to provide the dipole moment components.
Since the derived expected intensity ratio relies on the total dipole moment, it remains the same for optimized geometry.
For the computation of the dipole moment components, we use various levels of theory. 
Among them, our calculations at the HF level, yielded excellent
agreement with the existing experimental results. 
\cite{laka03} already analyzed permanent electric dipole moments of some 
aliphatic primary amines. They used various models for comparing 
their calculated results
with the experimentally obtained results. They found that HF/6-31G(3df)
model is more reliable for the aliphatic amines. 
According to their calculations, on an average, this 
model can predict values of permanent electric dipole moments with a deviation of only 
2.1\% of its experimental values. With reference to their results, here, 
we use the same method and the basis set to compute the dipole moment components. 
In Table 3, we show our calculated dipole moment components along with the experimental 
values, whenever available. 
Table 3 depicts that for most of the cases our estimated total dipole moments 
are in good agreement with the experimentally available data. In case of E-Ethanimine of 
$\rm{C_2H_5N}$ isomeric group, we found a maximum deviation of ($11.5$\%) between our calculated 
and experimental values of total dipole moments. On an average, we found a $5.35\%$ deviation
between our calculated and experimental.

As mentioned earlier, the rotational spectroscopy is the most convenient as
well as the most reliable method for detecting molecules in the ISM. Quantum
chemical studies have succeeded in providing reliable spectroscopic constants
to aid laboratory microwave studies.
Accurate quantum chemical studies of rotational transition frequencies may lead to interstellar
detections with confidence. Calculations of the rotational level need high-level basis sets
for accurate estimations of structure, spectrum and for optimization to obtain the
ground state energy. We use MP2 perturbation method with 6-311++G(d,p) basis set
which is capable of producing spectroscopic constants close to the experimental values.
Corrections for the interaction between rotational motion and vibrational motion along with 
corrections for vibrational averaging
and anharmonic corrections to the vibrational motion are also considered in our calculations.
In Table 4, we summarize our calculated theoretical values of rotational constants for 
all the species considered here. A comparison with the existing experimental results, whenever available 
is also made.  These spectroscopic constants can be used to generate catalog 
files of spectroscopic frequencies by using the SPCAT program \citep{pick91}
in the JPL/CDMS format. Table 4 also contains the rotational partition function
of a temperature relevant to the hot core condition ($\sim 200$ K). 
Among all the species considered here, $\lambda^1$-azanylmethane is a 
prolate symmetric top and trimethylamine is an oblate symmetric top 
and both have $3$ rotational symmetries.  Rest of the species in this study 
are asymmetric top having rotational symmetry $1$. We calculate the rotational partition function for 
the asymmetric top species by,
$$
Q_{rot}=5.3311 \times 10^6 \sqrt(T^3/ABC)/\sigma,
$$
where, $\sigma$ is the rotational symmetry number. Rotational partition function for the 
prolate symmetric top molecule is calculated by,
$$
Q_{rot}=5.3311 \times 10^6 \sqrt(T^3/B^2A)/\sigma.
$$
For the oblate symmetric top molecule, rotational partition function is calculated as, 
$$ Q_{rot}=5.3311 \times 10^6 \sqrt(T^3/A^2C)/\sigma.$$

\subsection{Chemical modeling}
Our large gas-grain chemical model \citep{das08b,das13a,das13b,maju14a,maju14b,das15a,das15b,
gora17a,gora17b}
is employed for the purpose of chemical modeling.
We assume that  gas and grains are coupled through accretion and thermal/non-thermal
desorption. Unless otherwise stated, a moderate value of the non-thermal desorption factor 
$\sim 0.03$ is assumed 
as mentioned in \cite{garr07}. 
A visual extinction of  $150$ and a cosmic ray ionization rate of 
$1.3\times 10^{-17}$ $\rm s^{-1}$ are used. 
The initial condition is adopted from \cite{leun84}. In order to mimic
actual physical conditions of the star-forming region, we consider the warm-up method that 
was established by \cite{garr06}.
Initially, we assume that the cloud remains in isothermal ($T=10$ K) phase for
$10^6$ years which is then followed by a subsequent warm-up phase where the temperature can
gradually increase up to $200$ K in $10^5$ years. 
So, our simulation time is restricted 
to $1.1 \times 10^6$ years. We assume that each phase has the same constant density ($n_H=10^7$ 
cm$^{-3}$).

Our gas phase chemical network is principally adopted from the UMIST 2012 
database \citep{mcel13}. For the grain surface reaction network, we primarily follow \cite{ruau16}. 
In addition to the above network, our network includes some reactions which are needed for the 
formation/destruction of interstellar amines and aldimines.
Ice phase formation of some of these amines and aldimines which are considered here are shown in Table 1. Similar pathways are also considered for the formation of these species in the gas phase 
as well.

For the computation of the gas phase rate coefficients of some 
additional gas phase neutral-radical (NR) reactions with barrier, we use transition 
state theory (TST), which leads to the Eyring equation \citep{eyri35}:
\begin{equation}
        k= (K_BT/hc) \exp(-\Delta {G} \ddag/RT) \ s^{-1},
\end{equation}
where, $\Delta {G}\ddag$ is the  Gibbs free energy of activation and `c' is the concentration 
which is 
set to $1$. $\Delta {G}\ddag$ is calculated by the quantum chemical calculation 
(QST2 method with B3LYP/6-311++G(d,p)).  Equation 1 depicts that rate
coefficient is exponentially increasing with the temperature. Thus, to avoid any
unattainable rate coefficient around the high temperature domain, we use an upper limit 
($10^{-10}$ cm$^3$s$^{-1}$) for Eqn. 1.

Normally, radical-radical addition reaction with a single product can occur 
through the radiative association. 
\cite{vasy13} outlined the rate coefficient for the formation of larger molecules 
by gas phase radiative association reactions. According to them, a larger molecule such as,
$\mathrm{CH_3OCH_3}$ can be formed by,
$$
\mathrm{CH_3+ CH_3O\rightarrow CH_3OCH_3 + Photon}.
$$
They considered the following temperature dependent rate coefficient for the above reaction: 
$$
k = 10^{-15} (T/300)^{-3}.
$$
In our work, we also consider similar rate coefficients for the radical-radical 
gas phase reactions leading to a single product.
In our model, we consider the formation and destruction of these
species in both the phases.

To compute the rate coefficients of ice phase reaction pathways, we use
diffusive reactions with a barrier against diffusion ($\kappa \times R_{diff}$) which is
based on thermal diffusion \citep{hase92}. $\kappa$ is the quantum mechanical probability 
of tunneling through a rectangular barrier of thickness $d$.
$\kappa$ is unity in the absence of a barrier. For reactions with activation energy
barriers ($E_a$), $\kappa$ is defined as the quantum mechanical probability for
tunneling through the rectangular barrier of thickness $d \ (= 1 \AA$) and 
is calculated by
\begin{equation}
\kappa= \exp [-2(d/ \hbar)(2\mu E_a)^{1/2}].
\end{equation}

Chemical enrichment of interstellar grain mantles depends on the 
desorption energies ($E_d$)
and  barriers against diffusion ($E_b$) of the adsorbed species. 
In the low-temperature regime,
the mobility of the lighter species such as, $\mathrm{H, \ D, \ N}$ and $\mathrm{O}$ mainly
controls the chemical composition of interstellar grain mantle. 
Composition of the grain mantle under the low-temperature regime is already discussed on several occasions \citep{chak06a,chak06b,das08a,das10,das11,sahu15,das16,sil17}.
Here, we use $E_b=0.50 E_d$ \citep{garr13}. 
Binding energies are mostly taken from KIDA databse.
Binding energy of some of the newly 
added ice phase species were not available in KIDA database. For these species, 
we have added the binding energies of the reactants which are required towards the
formation of these species. A similar technique was also employed in \cite{garr13}.
For example, for the calculation of the binding energy of $\rm{CH_3CNH}$, we add the
binding energies of $\rm{CH_3CN}$ and H.

For the destruction of gaseous amines and aldimines, we assume various ion-neutral (IN) and 
photo-dissociative pathways. Various IN and photo-dissociative destruction pathways were
already available in \cite{quan16} (for ethanimine) and \cite{mcel13} (for methanimine). We follow similar pathways and the
same rate coefficients for the destructions of other amines, aldimines and their associated species.
In the Appendix, we point out all the gas phase formation and destruction reactions which are
considered here. In analogy, for the destruction of ice phase amines, aldimines and their associated 
neutrals, we assume similar photo-dissociative reactions. Rate coefficients for the photo dissociative
reactions are assumed to be the same in both the phases. 
Abundances of the gas phase species can also decrease via 
adsorption onto the ice. However, the reverse process of desorption also occurs.  

\section{Results and Discussions}
In this Section, the results of high-level quantum chemical calculations together with 
our chemical model are presented and discussed. A detailed discussion of each isomeric 
group is given below.

\subsection{${CH_3N}$ Isomeric Group}
This group contains two molecular species (Figure 1), methanimine and $\lambda^1$-azanylmethane.
Methanimine ($\mathrm{CH_2NH}$) has already been observed long ago using Parkes $64$ m 
telescope in Sgr B2 \citep{godf73}. However, the presence of $\lambda^1$-azanylmethane is 
yet to be ascertained. 
Based on the enthalpy of formation and relative energy values shown in Table 2
methanimine appears to be the most stable candidate of $\rm{CH_3N}$ isomeric group. 
But enthalpy of formation is not sufficient enough
to dictate the abundance of this species
specifically when the system is far away from the equilibrium.
It is only the reaction pathways which can dictate the final abundance of any species in the ISM.
Our calculated dipole moment components (shown in Table 3) of methanimine are very close 
to the available experimental values. From our calculated dipole moment components, 
it is found that for methanimine, `a' and `b' type rotational 
transitions are the strongest whereas `c' type transitions are absent. 
In the case of $\lambda^1$-azanylmethane, the strongest component of dipole moment is 
found to be the `a' component whereas the `b' component is found to be the weakest.
The average dipole moment component of methanimine is found to be slightly higher than that of the
$\lambda^1$-azanylmethane. 
Our calculated
rotational constants for methanimine are also shown in Table 4 which are found to be very close to 
the prevailing experimental values.

It is believed that the methanimine is primarily created within the cold ice phase. The dominated 
pathways are shown in the reaction range R4-R7 of Table 1. Starting with the 
cyanide radical, $\rm{CH_2NH}$ may form through the successive hydrogen addition reaction in ice phase.
Subsequent hydrogen addition may take place in two ways: hydrogen addition with HCN could
result in $\rm{H_2CN}$ (R4) or HCNH (R5). \cite{woon02} pointed out that
reactions R4 and R5 possess activation energy barriers of about 
$3647$ K and $6440$ K respectively. $\rm{H_2CN}$ and 
HCNH can further produce $\rm{CH_2NH}$ by the hydrogen addition reaction (R6 and R7 respectively).
Surface network of KIDA already considered the reactions enlisted in \cite{gran14} and thus 
HCN/HNC related chemistry is consistent. 
The gas phase
pathways of \cite{gran14} is also considered in our gas phase network.
Near the higher temperatures, methanimine may be produced by the decomposition of
methylamine ($\mathrm{CH_3NH_2}$) \citep{john72}.
Recently, \cite{suzu16} pointed out that this species could be produced on the interstellar ice 
by other reactions (R1-R3) shown in Table 1.

For the gas phase reactions
G4 and G5 of Table 5, we obtain $\Delta G\ddag$ to be $8.37$ and $10.06$ Kcal/mol respectively. 
In Figure 2ab, chemical evolution of methanimine within the cold isothermal phase and in Figure 3 the 
subsequent warm-up phase is shown.
Abundances are shown with respect to $\rm{H_2}$ molecules.
It is clear that during the isothermal phase, methanimine is significantly abundant in both the phases
and having a peak abundance of  $5.93 \times 10^{-08}$ in gas phase and 
$3.49 \times 10^{-06}$ in the ice phase. Strong decreasing slope of gas phase methanimine is
observed from Figure 2a. Figure 2b also depicts a decreasing slope (during the end of the isothermal regime) of ice phase methanimine due to the production of methylamine by successive hydrogen addition reaction (R8-R9).
Dashed curves in Figure 2a are shown for the gas phase abundances of all the 
aldimines and amines for the case where non-thermal desorption factor, $a_{fac}$ is assumed to be $0$. 
The gas phase abundance of methanimine with 
$a_{fac}=0$ (dashed line in Figure 2a) and $a_{fac}=0.03$ (solid line in Figure 2a) 
differ significantly. 
It is because in the isothermal phase, the gas phase contribution of methanimine is mainly coming 
from the ice phase via non-thermal desorption mechanism. Following KIDA database, desorption energy 
of methanimine is assumed to be $5534$ K.
It is evident that in the warm-up phase sublimation of methanimine occurs around
$110$K. In the warm-up phase (Figure 3), abundance of gas phase methanimine is significantly 
increased due to the efficient gas phase formation by reactions G1-G7.
Peak abundance of gas phase methanimine is found to be around 
$1.81 \times 10^{-09}$. Our obtained values can be compared with the 
recent hot-core observation of methanimine 
$\sim 7.0 \times 10^{-08}$ in G10.47+0.03 and $3.0 \times 10^{-9}$ in NGC6334F by \cite{suzu16}. 

\begin{figure}
\centering
\includegraphics[width=0.35\textwidth]{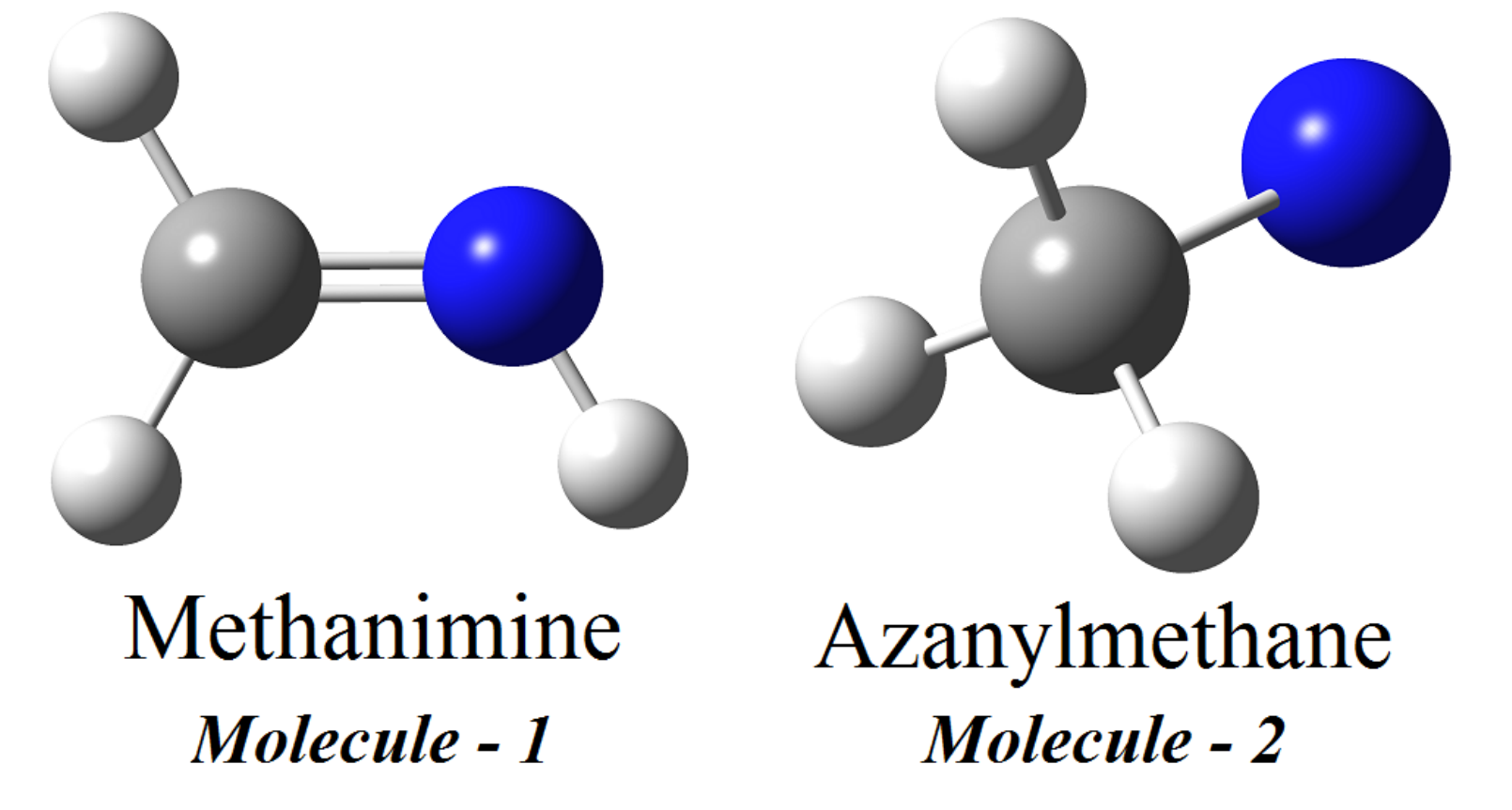}
\caption{$\mathrm{CH_3N}$ isomers.}
\end{figure}

\begin{figure*}
\centering
\includegraphics[height=9cm, width=11cm]{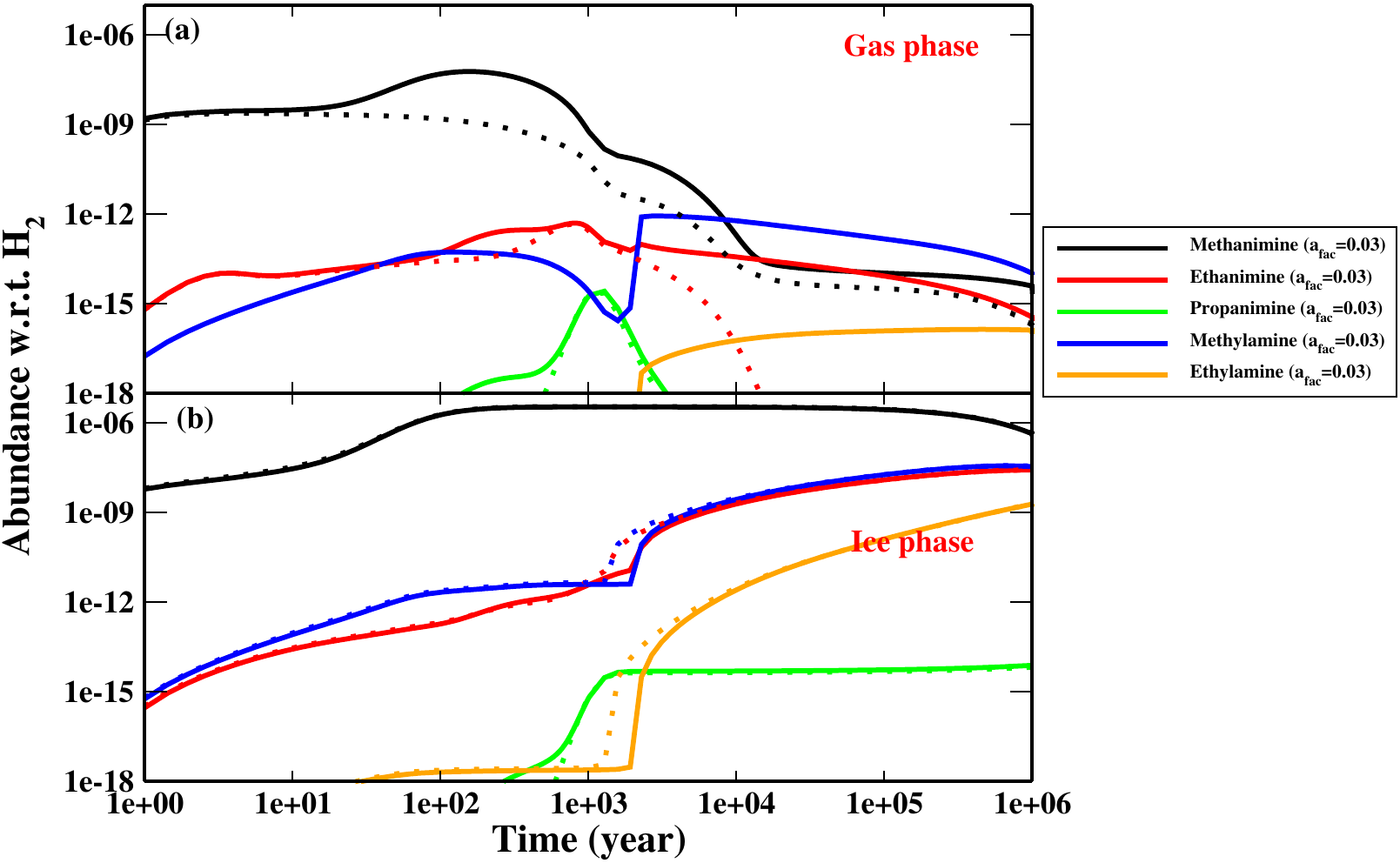}
\caption{Chemical evolution of the aldimines and amines in the isothermal phase for $a_{fac}=0.03$ (solid) and 
$0$ (dashed).}
\end{figure*}

\begin{figure*}
\centering
\includegraphics[height=9cm, width=12cm]{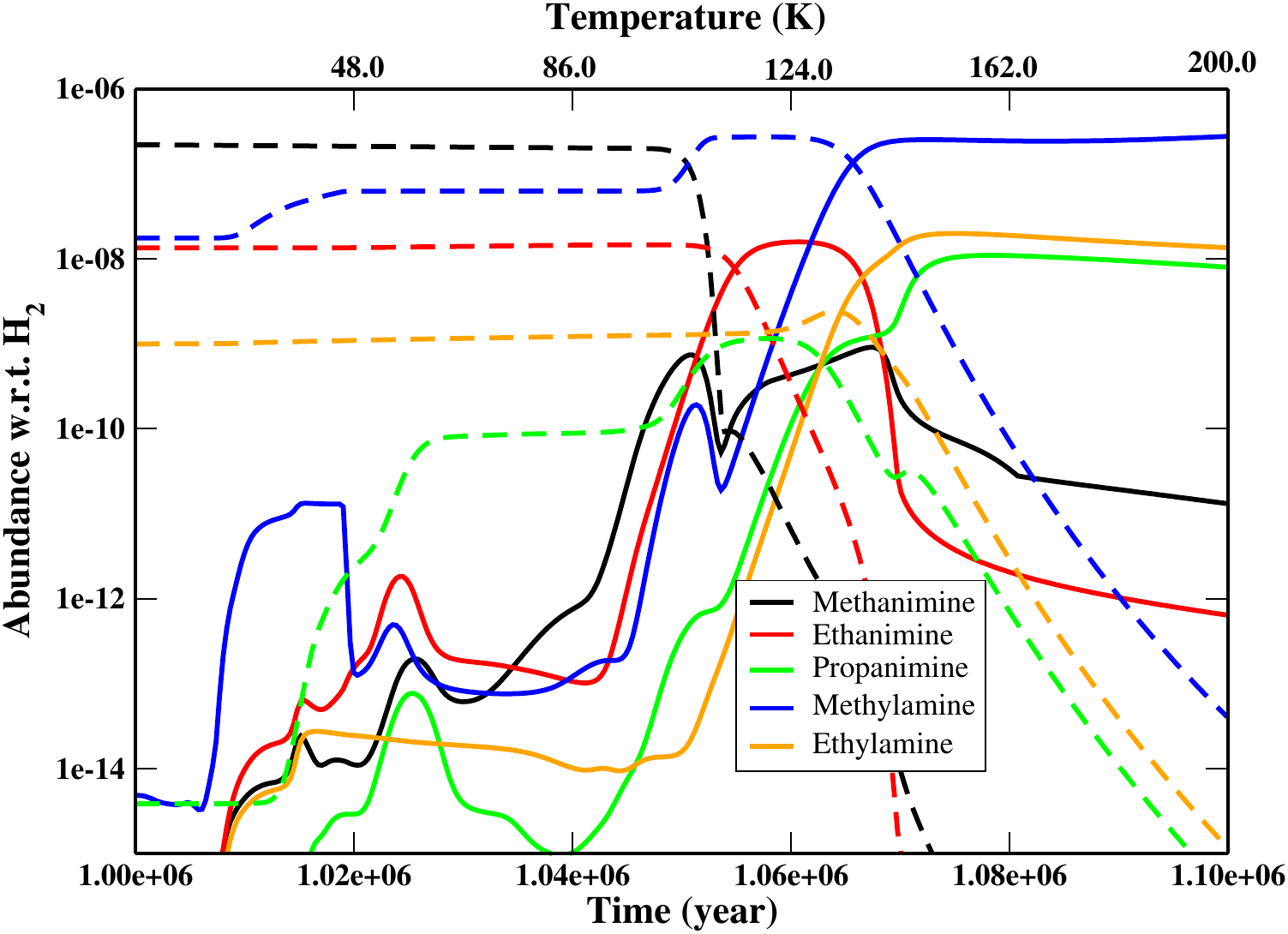}
\caption{Chemical evolution of the aldimines and amines in the warm-up phase. 
Solid lines represent the gas phase species, whereas corresponding dashed lines represent the ice phase species.}
\end{figure*}

\subsection{${CH_5N}$ Isomeric Group}
Only methylamine ($\mathrm{CH_3NH_2}$) belongs to this isomeric group (Figure 4) and this was
already observed long ago in Sgr B2 and Ori A \citep{kaif74,four74}.  
In Table 2, we compare our calculated enthalpies of formation with that of the existing experimental
value. We find that our calculated value with the B3LYP/6-31G(d,p) method is closer to the 
experimentally obtained value than that computed from the G4 composite method.
Methylamine is the precursor 
of an amino acid (glycine) formation. \cite{taka73,kaif74} found that the c-type transitions 
of methylamine are $4$ times stronger than the a-type transitions. 
We also found a very strong c-component of dipole moment
shown in Table 3. Calculated total dipole moment component for methylamine is $1.2874$ Debye
whereas the experimentally obtained value is $1.31 \pm 0.03$ Debye (Table 3).
Also, a very good correlation between our calculated rotational constants and experimentally obtained 
values can be seen from Table 4.

\begin{figure}
\centering
\includegraphics[width=0.35\textwidth]{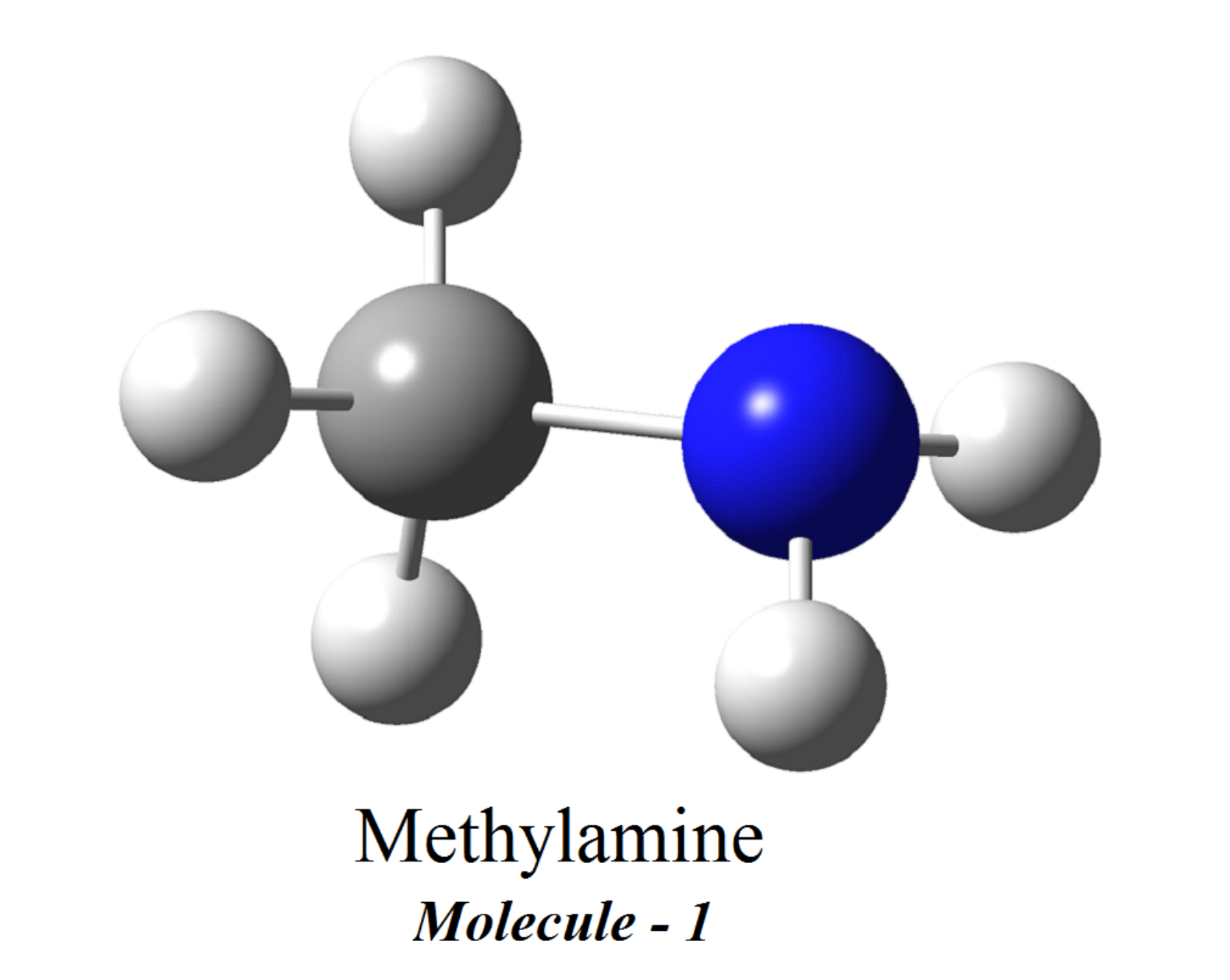}
\caption{$\mathrm{CH_5N}$ isomer.}
\end{figure}

In the ISM, methylamine may be formed via two successive 
hydrogen addition reactions of methanimine in the ice phase \citep{godf73,suzu16}. 
\cite{woon02} determined that the primary step of this hydrogen addition reaction may proceed in 
two ways. First, hydrogenation of methanimine yields $\rm{CH_3NH}$ (R8) having activation 
barrier of $2134$ K and second, it may produce $\rm{CH_2NH_2}$ (R9) having activation barrier of $3170$ K.
Reaction R8 and R9 may also occur in the gas phase as well. From our TST calculation, we have
$\Delta G\ddag= 7.84$ Kcal/mol for the reaction G8 of Table 5. However,
we did not find a suitable transition state for the gas phase reaction G9 of Table 5. 
In the ice phase, R9 possesses higher activation barrier ($1.485$ times higher) than R8.
We assume that a similar trend would be followed for the gas phase reaction G9 and so, we assume
{$\Delta G\ddag=11.64$} Kcal/mol for the gas phase reaction G9.
Methylamine may further be produced by hydrogenation reaction of these two products by
reaction numbers R10 and R11 respectively. 
\cite{woon02} recommended that the simplest amino acid, glycine may be formed by the reaction 
between $\mathrm{CH_2NH_2}$ and COOH radical. So, methylamine is an important product
towards the formation of glycine.

Since reactions G8 and G9 possess high $\Delta G \ddag$, during the isothermal phase, production of
gas phase methylamine is inadequate. However, despite high activation barrier ($E_{\rm a}$) 
reactions R8 and R9 would be efficiently processed on interstellar ice by quantum mechanical tunneling and
populate the gas phase by the non-thermal desorption.
Mainly due to the non-thermal chemical desorption phenomenon, ice phase methylamine populate the gas phase.
It is clearly visible from Figure 2a 
that for the case of $a_{fac}=0$, 
gas phase contribution of methylamine is negligible (the dashed line corresponding 
to the methylamine is absent in Figure 2a), 
So, in the isothermal phase, contribution for the 
gas phase methylamine is mainly comes from the ice phase.
We find that in the isothermal phase, methylamine attains a peak value of $8.46 \times 10^{-13}$ in the gas phase
and  $3.70 \times 10^{-08}$ in ice phase respectively. 
From Figure 3, it is observed that ice phase abundance initially 
increases due to the increase in the mobility of the reactants.
Peak ice phase abundance of methylamine is obtained to be  $5.44 \times 10^{-07}$.
Near the high temperature, production of gas phase methylamine significantly 
contributed due to (a) the enhancement of the temperature dependent rate coefficient of reactions 
G8 and G9 and (b) increase in the gas phase methanimine formation.
The peak gas phase abundance of methylamine is found 
to be  $5.54 \times 10^{-07}$. 
Our obtained values may be compared with the recent observation of methylamine \citep{ohis17}.
They predicted methylamine abundance $\sim 1.2 \times 10^{-08}$ in G10.47+0.03.

\subsection{${C_2H_5N}$ Isomeric Group}
Five isomers belong to this isomeric group (Figure 5): E-ethanimine, Z-ethanimine, ethenamine, 
N-methylmethanimine and aziridine. Out of these five isomers, 
recently, both the conformers (E and Z) of ethanimine ($\mathrm{CH_3CHNH}$) 
had been detected in Sgr B2 \citep{loom13}. From our quantum chemical calculation, we found that E-ethanimine is 
energetically more stable ($4.35$ KJ/mol by using MP2/6-311G++(d,p) and $1.2$ KJ/mol by using G4 composite method) 
than Z-ethanimine. \cite{quan16} and \cite{loom13} obtained 
an energy difference of $4.60$ KJ/mol and $4.24$ KJ/mol respectively between these two conformers. 
We have shown the enthalpy of formation values for all the species in Table 2 along with 
the experimentally obtained values, where available.  Our calculated 
enthalpies of formation using B3LYP/6-31G(d,p) method are in good agreement with the experimentally obtained values  
of E-ethanimine,  n-methylmethanimine and  Aziridine. 
For a better assessment, in Figure 6, we show the enthalpy of formation with  the molecule
number noted in Table 2 and Figure 5 for $\rm{C_2H_5N}$ isomeric group. Clearly, E-ethanimine has 
the minimum enthalpy of formation followed by Z-ethanimine. The energy difference between 
these two is smaller than the other isomers of this isomeric group. 
Observed isomers are marked as green circles and the unobserved 
are marked by red circles in Figure 6.

All the dipole moment components are presented in Table 3.
Our calculated values of dipole moments are compared 
with the existing values \citep{lova80,lias05}. The effective dipole moment of Z-ethanimine is found to be higher than that of E-ethanimine. \cite{char95} pointed out that for an optically thin emission, 
an idea about the antenna temperature could be made by calculating the intensity of a 
given rotational transition. 
This intensity is proportional to $\mathrm{\mu^2/Q(T_{rot})}$, where $\mu$ is the electric 
dipole moment and $\mathrm{Q (T_{rot})}$ is the partition function at rotational temperature 
($\mathrm{T_{rot}}$). In Table 4, we have pointed out the rotational constants and the
rotational partition function at $T_{rot}=200$ K. In Figure 7, we have shown the plot of the expected
intensity ratio (by assuming all the species of this isomeric group have same abundances) 
with respect to the  most stable isomer (E-ethanimine) of this isomeric 
group (relative energy values 
with molecule number of all isomer of $\mathrm {C_2H_5N}$ isomeric group noted in Table 2).
In Figure 7, the expected intensity ratios for all the species of this isomeric group 
are shown by considering the effective dipole moments.
Since in this isomeric group, Z-ethanimine has the largest effective dipole moment, assuming the same
abundances, probability of detecting Z-isomer of ethanimine will be more favourable than the
other isomers of this isomeric group. From Figure 7, we can see that after E and Z isomers 
of ethanimine, aziridine has the strongest transition but due to its higher relative energy 
in comparison to E-ethanimine makes it less probable candidate (if reaction pathways do not 
influence at all) for astronomical detection.

\begin{figure}
\centering
\includegraphics[width=0.45\textwidth]{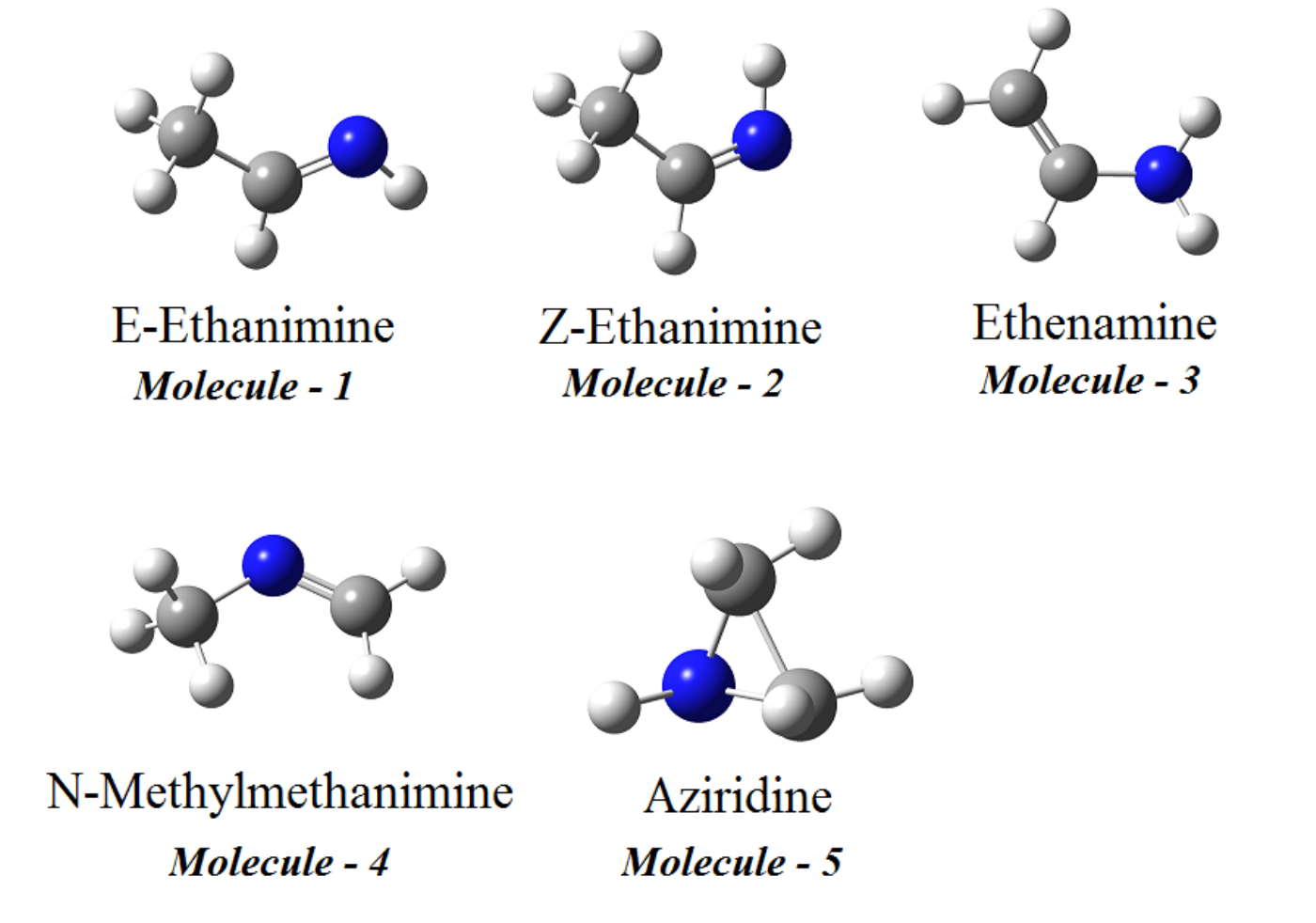}
\caption{$\mathrm{C_2H_5N}$ isomers.}
\end{figure}

\begin{figure}
\centering
\includegraphics[width=0.45\textwidth]{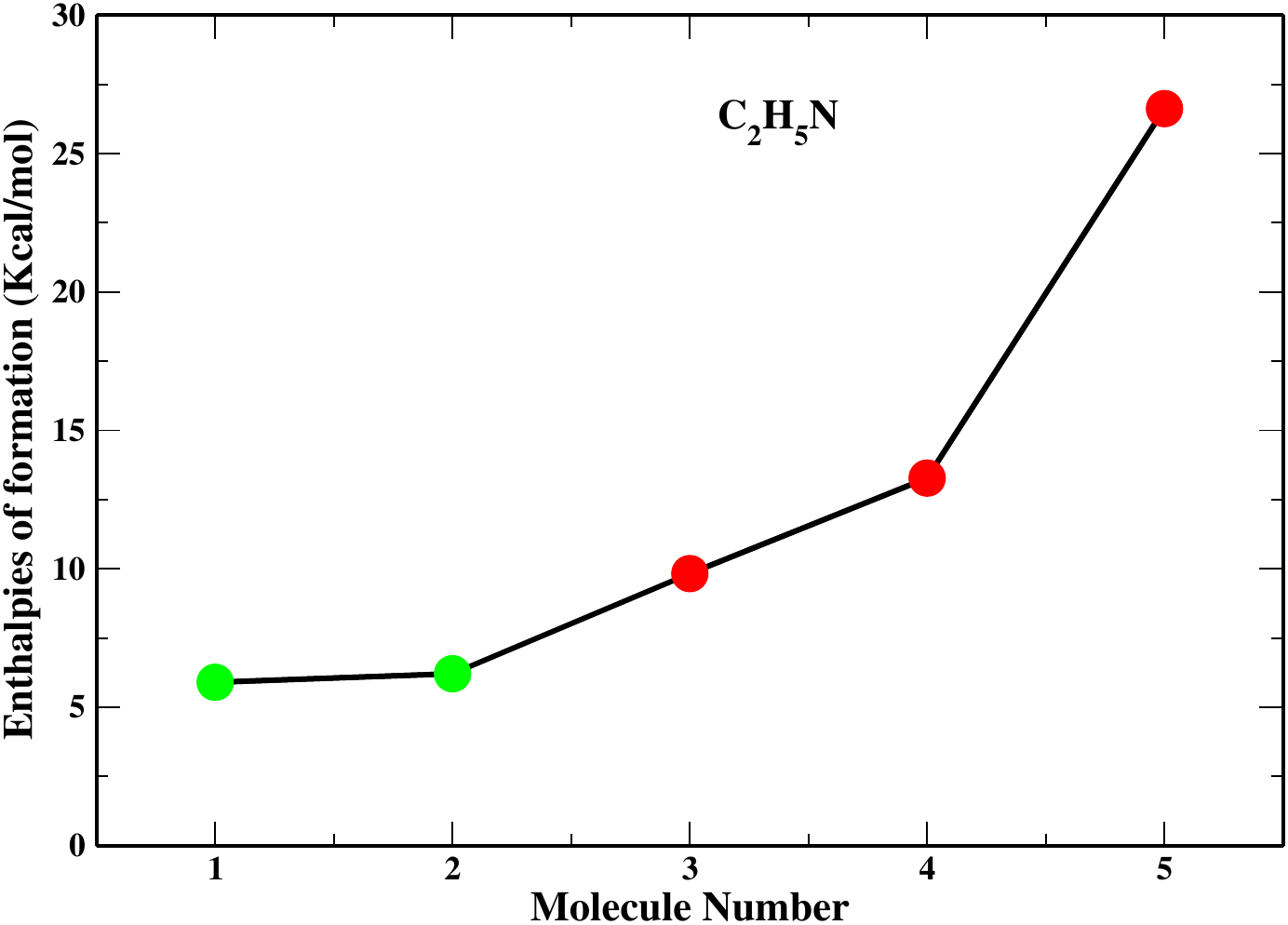}
\caption{Enthalpy of formation of $\rm{C_2H_5N}$ isomeric group. Molecules already observed are
marked as green circles and those yet to be observed are marked as red circles.}
\end{figure}

\begin{figure}
\centering
\includegraphics[width=0.45\textwidth]{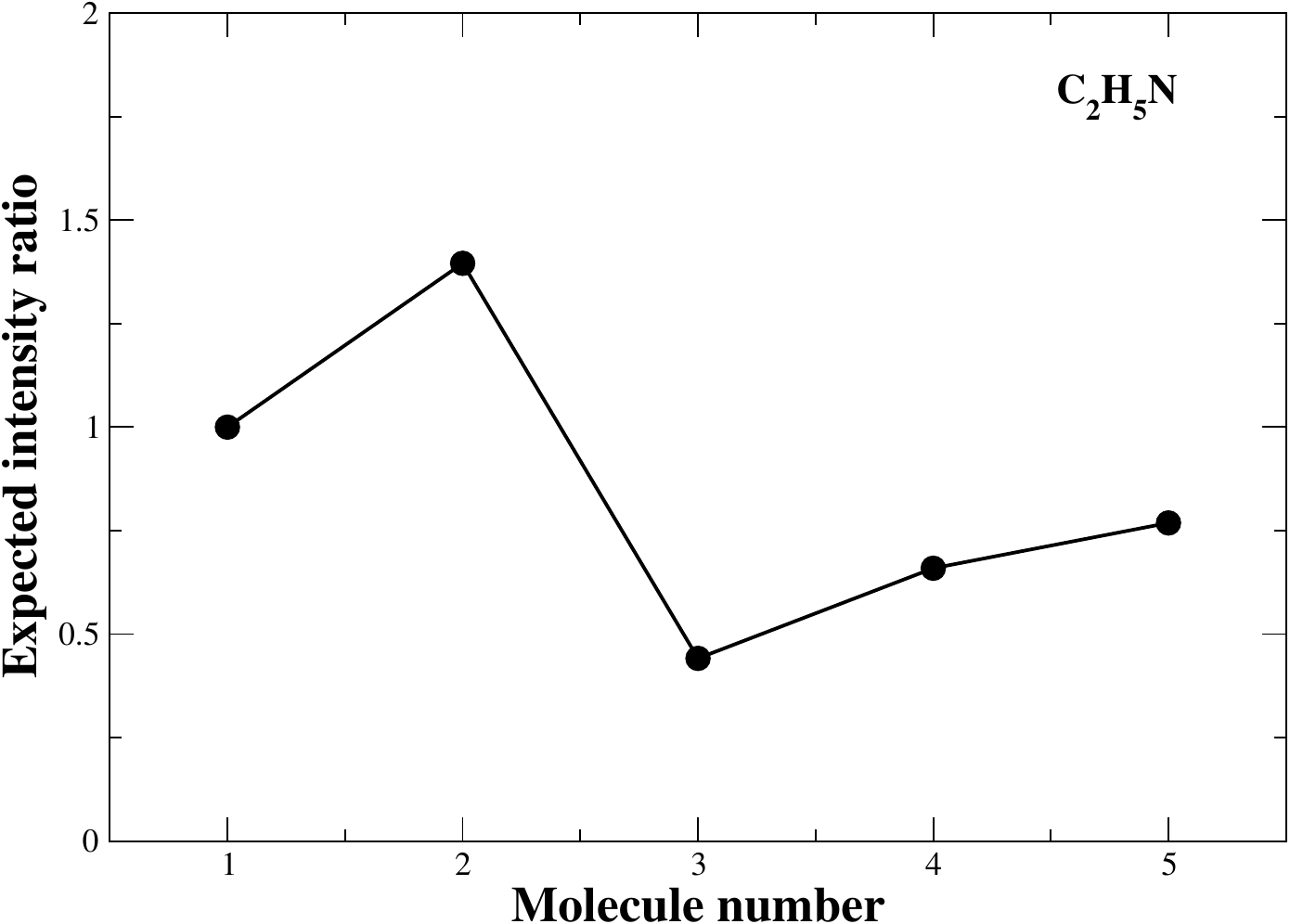}
\caption{Expected intensity ratio of $\rm{C_2H_5N}$ isomeric group by considering the effective dipole moment of all the species.}
\end{figure}

\cite{loom13} mentioned that in ice phase ethanimine may be produced via 
two consecutive hydrogen addition reactions with $\mathrm{CH_3CN}$. \cite{quan16} recently 
proposed that the first step (R14) hydrogen addition 
reaction with  $\mathrm{CH_3CN}$ has the barrier of $1400$ K and the second step (R15) is 
a radical-radical reaction and assumed to be barrier-less in nature. For the gas phase
reaction of G14 of Table 5, our calculated value of $\Delta G \ddag$ is $10.32$ Kcal/mol.
\cite{quan16} additionally suggested that ethanimine can even be produced by the reaction
between $\rm{CH_3}$ and $\rm{H_2CN}$ in ice phase. Among the gas phase pathways, reaction between
$\rm{C_2H_5}$ and NH (reaction G23 of Table 5) may lead to $\rm{CH_3CHNH}$. In our network, We include 
all the gas phase reactions mentioned in Table 5 and ice phase reactions shown in Table 1. 
For the gas phase barrier-less reactions, a conservative value of the rate coefficient
\citep{vasy13} are considered. Here, we assume only one form of ethanimine (E-ethanimine) for 
the purpose of our modeling.

In Figure 2ab, we have shown the chemical evolution of ethanimine 
in the isothermal phase and subsequently the warm-up phase is shown in Figure 3. In the isothermal phase, 
ethanimine has the peak value of 
$4.99 \times 10^{-13}$ in the gas phase and  $2.69 \times 10^{-08}$
in the ice phase respectively.
During the warm-up phase, gas phase ethanimine has a peak value of  
$3.17 \times 10^{-08}$. 
In the warm-up phase, gas phase production of ethanimine is also contributing due to the 
enhancement of the temperature dependent rate coefficient of reaction G14. 
Around $125$ K, abundance of ethanimine attains a peak and started to decrease due to the efficient production of ethylamine by the successive hydrogenation reactions (R17-R18).
Since the reaction R17 has a barrier, we use Eq.1 for the computation of its rate
coefficient. Eq. 1 clearly says that as we are increasing the temperature, rate 
coefficient increases exponentially. 
Thus, in the warm-up phase,
rate coefficient of reaction G17 increases 
exponentially and attain a reasonable rate ($10^{-10}$ cm$^3$ s$^{-1}$) which means the destruction 
of ethanimine by the hydrogenation reaction also gradually increases and attain a quasi-steady state.
At the end of our simulation (after $1.1 \times 10^6$ year), we are having 
an abundance of $1.27 \times 10^{-12}$ for Z-ethanimine whereas the
predicted abundance of Z-ethanimine $\sim 6.0 \times 10^{-11}$ by \cite{quan16}.

\subsection{${C_2H_7N}$ Isomeric Group} 
Trans-ethylamine, gauche-ethylamine and dimethylamine belong to the $\rm{C_2H_7N}$ isomeric group (Figure 8). 
Interestingly, no species of this isomeric group is yet to be detected in the ISM.
However,  the presence of ethylamine was traced in comet Wild 2  \citep{glav08}.
Ethylamine is the precursor of simple amino acid, glycine. It can exist in the form of 
two stable conformers: gauche and trans. Experiment by \cite{hama86} 
shows that the trans conformer is slightly more stable than the gauche conformer. Our calculated values are 
also in line with this result. We obtained that gauche conformer has $1.67$ KJ/mol 
higher energy than its trans conformer. 
So, according to the enthalpy of formation and relative energies are shown in Table 2, 
trans-ethylamine has the least enthalpy of formation and is most stable 
among this isomeric group. In Figure 9, enthalpy of 
formation of this isomeric group is depicted  with the molecule
number and enthalpy of formation is noted in Table 2.
In comparison with the experimentally measured enthalpies of formation, the G4 composite method 
overestimates the enthalpy of formation values for ethylamine and dimethylamine.

Based on the data available from our quantum chemical calculations, in Figure 10,
we have shown the expected intensity ratio with  respect to the species having least 
enthalpy of formation.  Figure 10 depicts that the trans-ethylamine has the highest expected
intensity ratio ($\sim 1$) in comparison with the other two members (gauche-ethylamine has $0.97$
and dimethylamine has $0.69$) of this isomeric group and thus
trans-ethylamine has the highest probability of its astronomical detection from this isomeric group.

Ethylamine could be formed on the grain surface via two successive hydrogen additions of 
ethanimine. Our calculation reveals that the first step of this hydrogenation reaction 
having activation barrier of $1846$ K (R17). For the gas phase hydrogenation reaction (G17),
our calculated $\Delta G \ddag$ parameter is found to be  $9.98$ Kcal/mol.
Since the second step of this
reaction is radical-radical in nature, we assume that reaction R18 may be treated as a 
barrier-less process. Gas and ice phase abundances of ethylamine in the isothermal phase 
are depicted in Figure 2ab. 
In the isothermal phase, we have a peak gas phase abundance of  $1.19 \times 10^{-16}$ 
and in the ice phase  $1.99 \times 10^{-09}$.
In the warm-up phase (Figure 3), ice phase abundance roughly remains invariant up to $125$ K
and then it starts to decrease sharply. Its gas phase abundance is 
higher and has a peak abundance of $3.98 \times 10^{-08}$. 

In terms of the size of the molecule, ethanimine is more complex than methanimine. Similarly,
their successors, ethylamine is more complex than methylamine. So, it may be expected that throughout
the evolutionary stage, the abundances of ethanimine/ethylamine would be always less than that 
of methanimine/methylamine. But Figure 2ab and Figure 3 depicts that this trend is not
universal for all circumstances. This is due to the fact that the formation of methanimine and 
its successor and ethanimine and its successor are processed through totally different channels.
Their destruction rates are also different. 
For example, ice phase formation of methanimine mainly occurs by successive hydrogenation reactions 
with HCN (reaction R4 to R7) whereas ethanimine formation is mainly controlled by 
successive hydrogenation reactions with CH$_3$CN (reaction R14 and R15). Now, reactions 
R4 and R5 contain much higher barrier than that of the reaction R14. Since HCN is more
abundant than CH$_3$CN, despite high barriers involved in the formation of methanimine,
in the isothermal stage, most of the time, ice phase abundance of methanimine remains  
higher than ethanimine. In the hot core region, due to lower activation barrier of reaction R14, 
ethanimine formation become more favourable. Now, methylamine is forming from methanimine 
(by reaction numbers R8-R10) and ethylamine is forming from ethanimine (by reaction numbers R17-R18).
Once again, the activation barrier involved in case of methylamine formation (reaction R9) is higher
than the barrier involved in the formation of ethylamine (reaction R17). Since, very
complex chemistry is going on, abundances of these species
should be compared very carefully.

\begin{figure}
\centering
\includegraphics[width=0.45\textwidth]{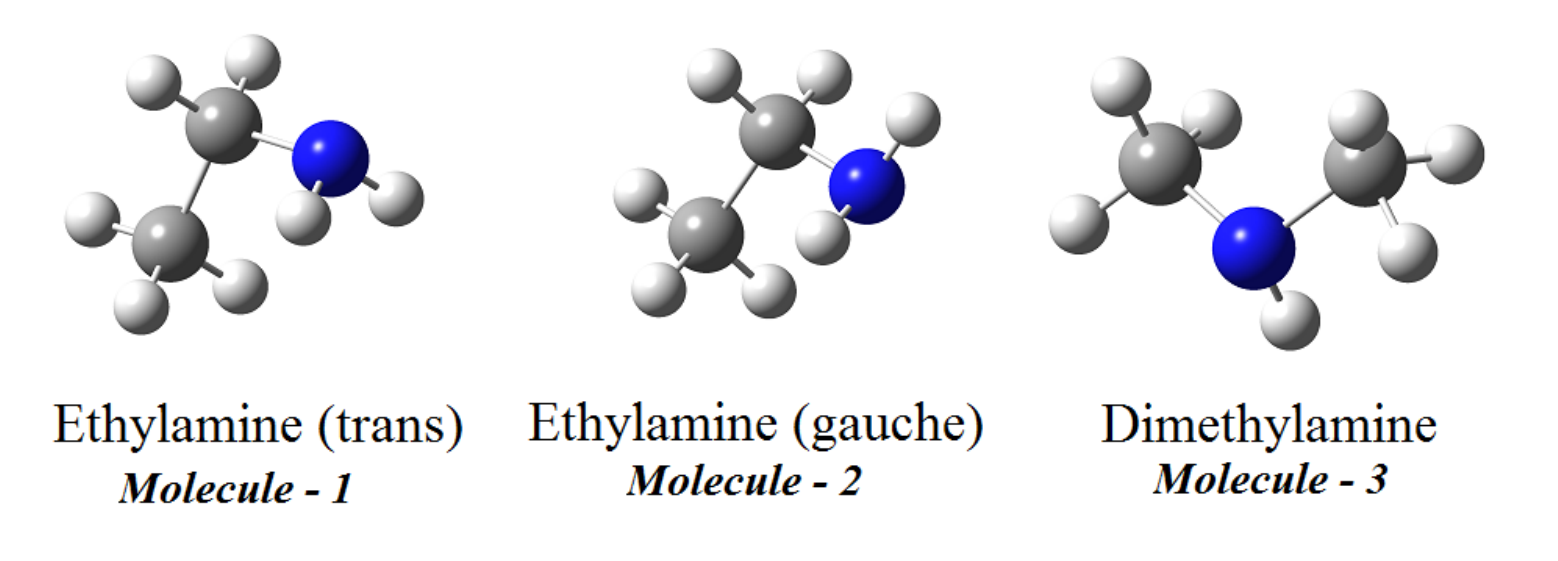}
\caption{$\mathrm{C_2H_7N}$ isomers.}
\end{figure}

\begin{figure}
\centering
\includegraphics[width=0.45\textwidth]{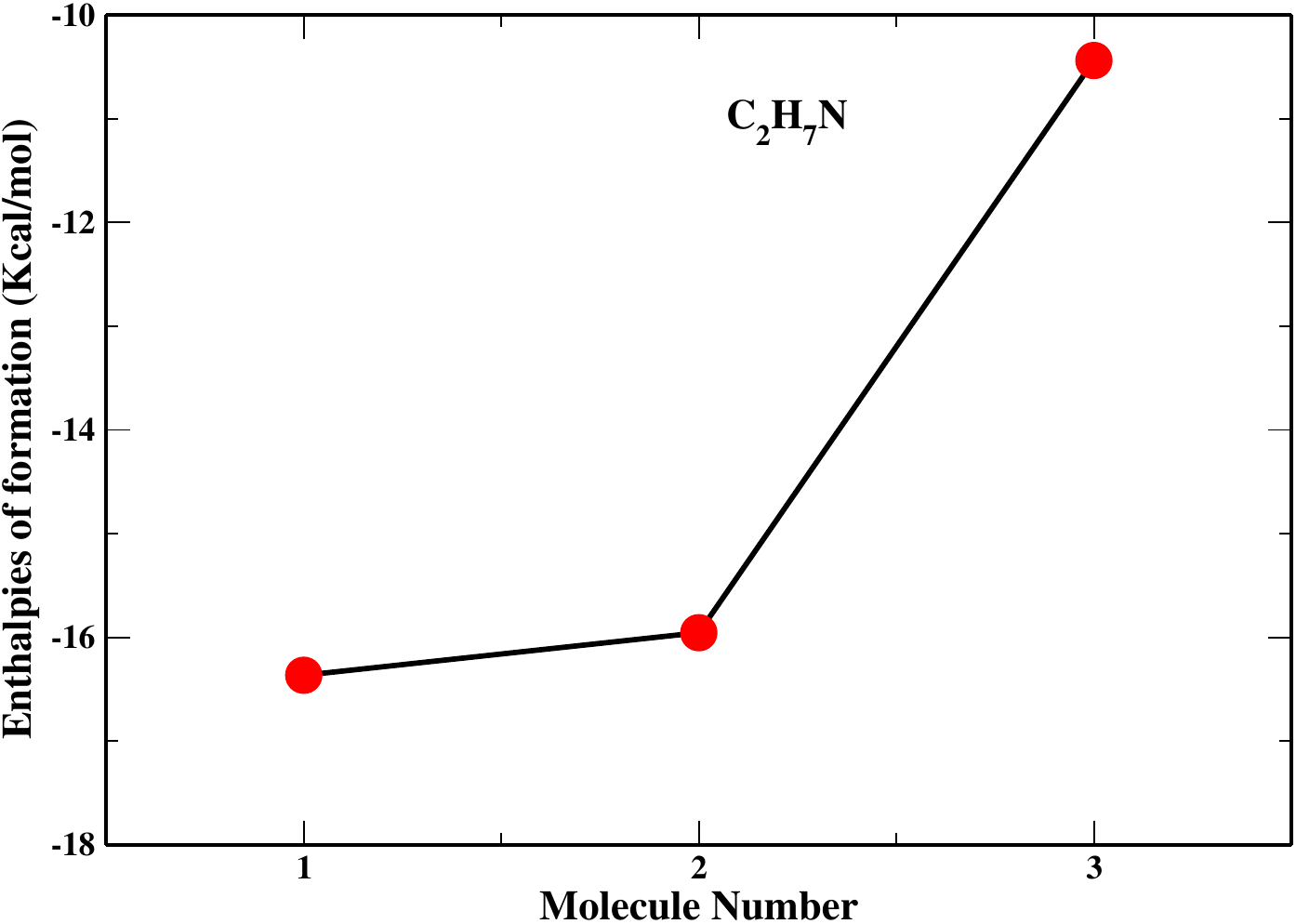}
\caption{Enthalpy of formation $\rm{C_2H_7N}$ isomeric group.}
\end{figure}

\begin{figure}
\centering
\includegraphics[width=0.45\textwidth]{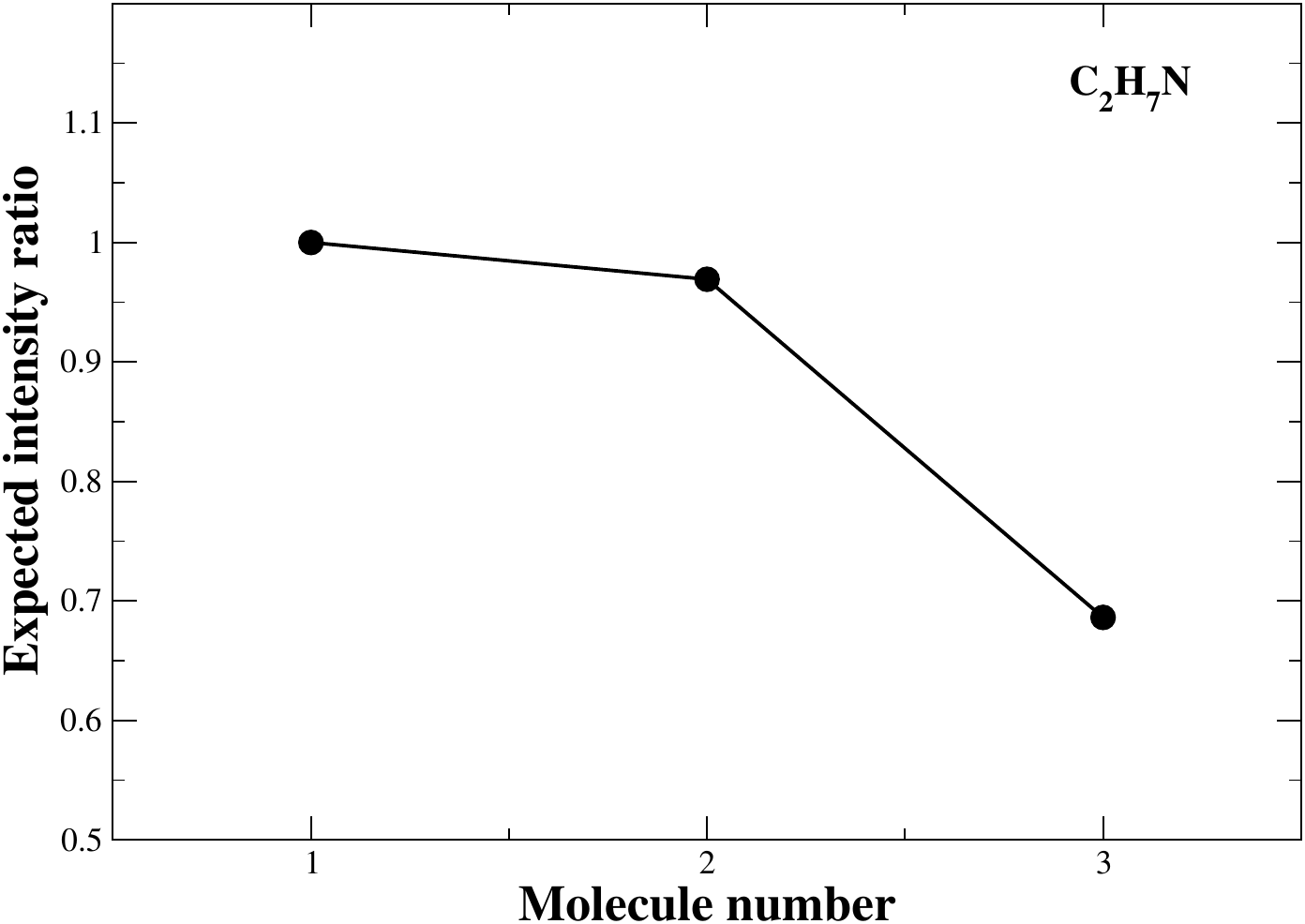}
\caption{Expected intensity ratio of $\rm{C_2H_7N}$ isomeric group by considering the effective dipole moment of all the species.}
\end{figure}

\cite{altw16} showed that in 67P/Churyumov-Gerasimenko, relative abundance between 
methylamine and glycine is $1.0\pm0.5$ and ethylamine to glycine ratio is $0.3 \pm0.2$. 
Taking the maximum and minimum values from this observation, we can see that 
methylamine to ethylamine ratio may vary in between $\sim 1-15$. 
In order to check the correlation (if any) between the cometary ice as observed by \cite{altw16} and 
interstellar ice, we may focus on our ice phase evolution results of the isothermal ($T=10$ K) phase.
From our modeling results (Figure 2b), we found that in the isothermal phase 
(at time 10$^6$ year), the methylamine to ethylamine ratio to be of $\sim 17.7$ in 
the ice phase which is very close to the observed value \citep{altw16}. 
This suggests that more
in depth study is required to confirm this linkage between the interstellar and cometary origin
of these molecules. 

\subsection{${C_3H_7N}$ Isomeric Group}

Nineteen isomers (Figure 11) belong to $\rm{C_3H_7N}$ isomeric group. Figure 12 depicts enthalpy 
of formation ($\mathrm{\Delta_fH^O}$) of this isomeric group. In Table 2, we
show the relative energies  and enthalpy of formation of 
various isomers of this isomeric group. 
Only for cyclopropanamine, experimentally obtained enthalpy of formation is available and 
this is in close agreement with our calculated enthalpies of formation with the B3LYP/6-31G(d,p) method.
Clearly, 2-propanimine is the most stable isomer of this group followed by
2-propenamine. Next species in this sequence is (1E)-1-propanimine. (1E)-1-propanimine has the lower energy 
than that of the (1Z)-1-propanimine.

Though (Dimethyliminio)methanide (molecule no. 19) has the highest enthalpy of formation and 
is the least stable of the species in this isomeric group, interestingly, our calculation 
listed in Table 4 shows that it possesses the highest effective electric dipole moment. 
(1Z)-1-propanimine (molecule no. 6) is found to be having the second 
highest effective dipole moment in this isomeric group. However, it is found that the a-type 
transitions of 1-Z propanimine are the strongest among all the 
species of $\rm{C_3H_7N}$ isomeric group. Figure 13, shows the expected intensity ratio (by considering the effective dipole moment) 
with respect to the most stable (as well as the species having
least enthalpy of formation) isomer. From Figure 13, it is clear that if the abundances of all 
these isomeric species are assumed to be the same, 
then (Dimethyliminio)methanide and (1Z)-1-propanimine may be the most probable 
candidate for the astronomical detection from this group. Since (Dimethyliminio)methanide is not
very stable species, it does not have high probability of detection. 
Thus, based on the stability, enthalpy of formation and expected intensity ratio, 
(1Z)-1-propanimine is the most suitable species for the future astronomical detection from this
isomeric group. However, it is the reaction pathways which can ultimately decide the
fate of this species.

Methanimine and ethanimine have already been observed in the ISM and (1Z)1-propanimine 
may be the next probable candidate for astronomical detection. To the best of our knowledge, the 
astronomical searches of (1Z)-1-propanimine are yet to be reported in the literature. 
So (1Z)-1-propanimine remains the best candidate for astronomical observation among all the 
isomers of $\mathrm{C_3H_7N}$ isomeric group.  

\begin{figure}
\centering
\includegraphics[width=0.45\textwidth]{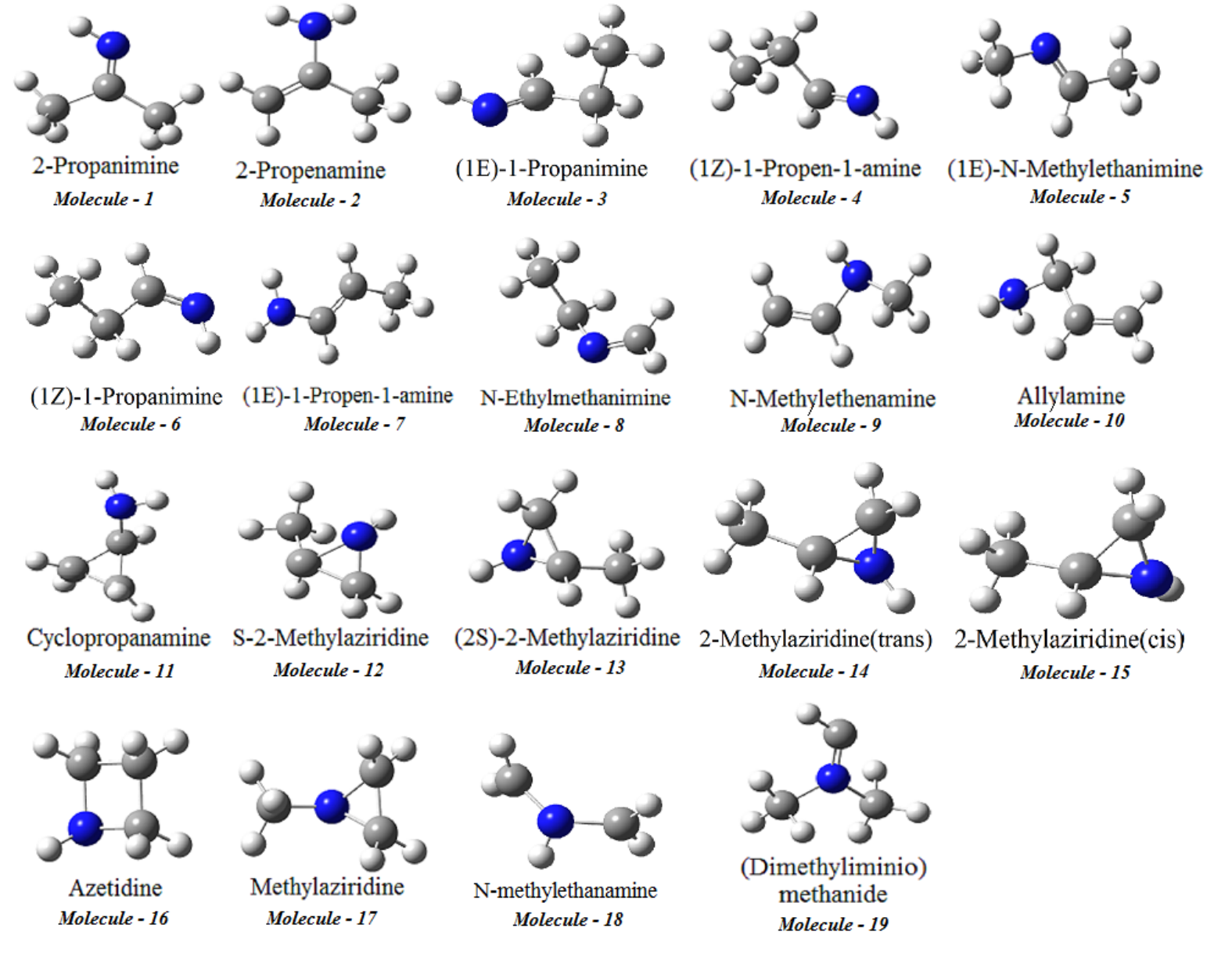}
\caption{$\rm{C_3H_7N}$ isomers.}
\end{figure}

\begin{figure}
\centering
\includegraphics[width=0.45\textwidth]{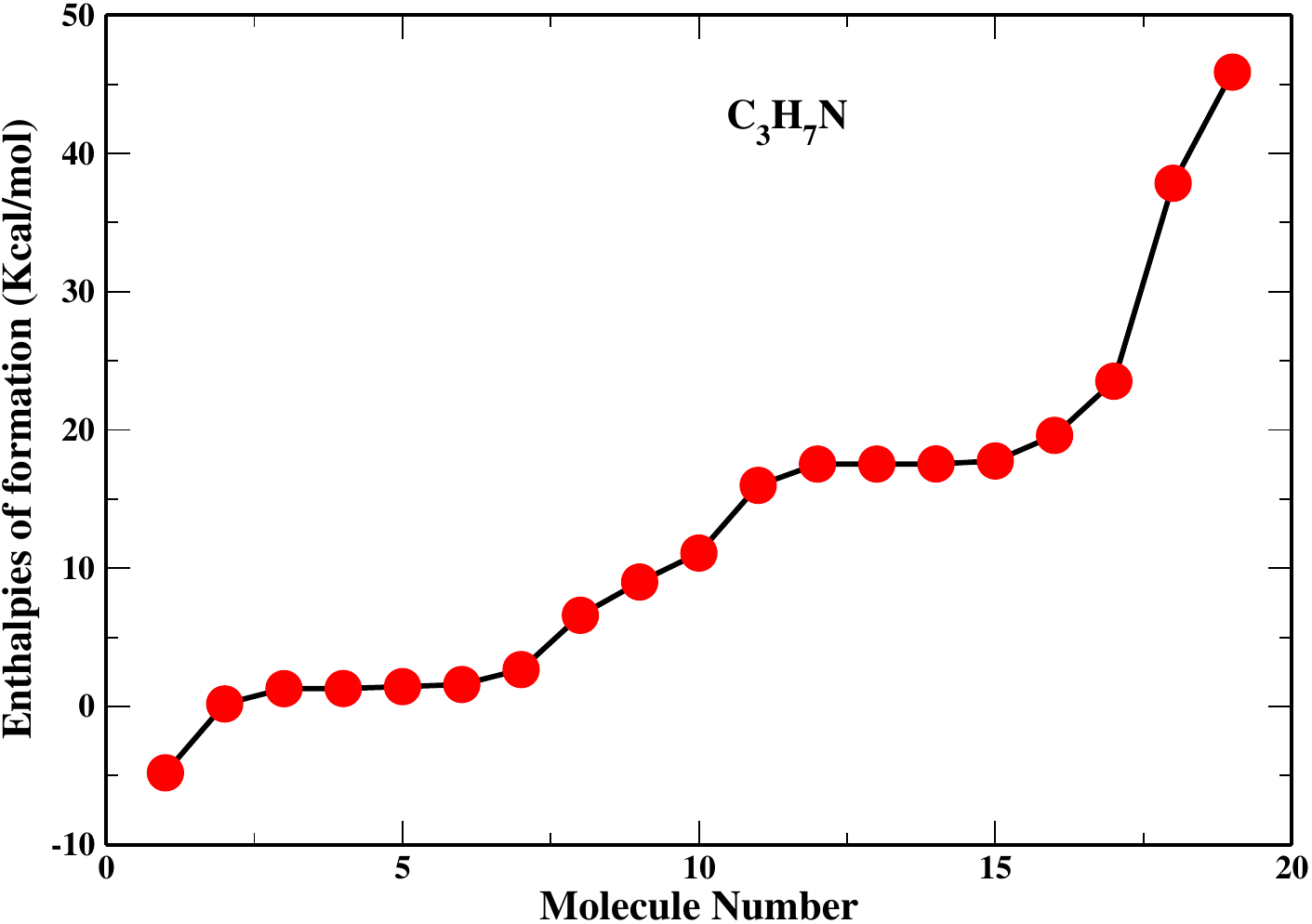}
\caption{Enthalpy of formation of $\rm{C_3H_7N}$ isomeric group.}
\end{figure}

\begin{figure}
\centering
\includegraphics[width=0.45\textwidth]{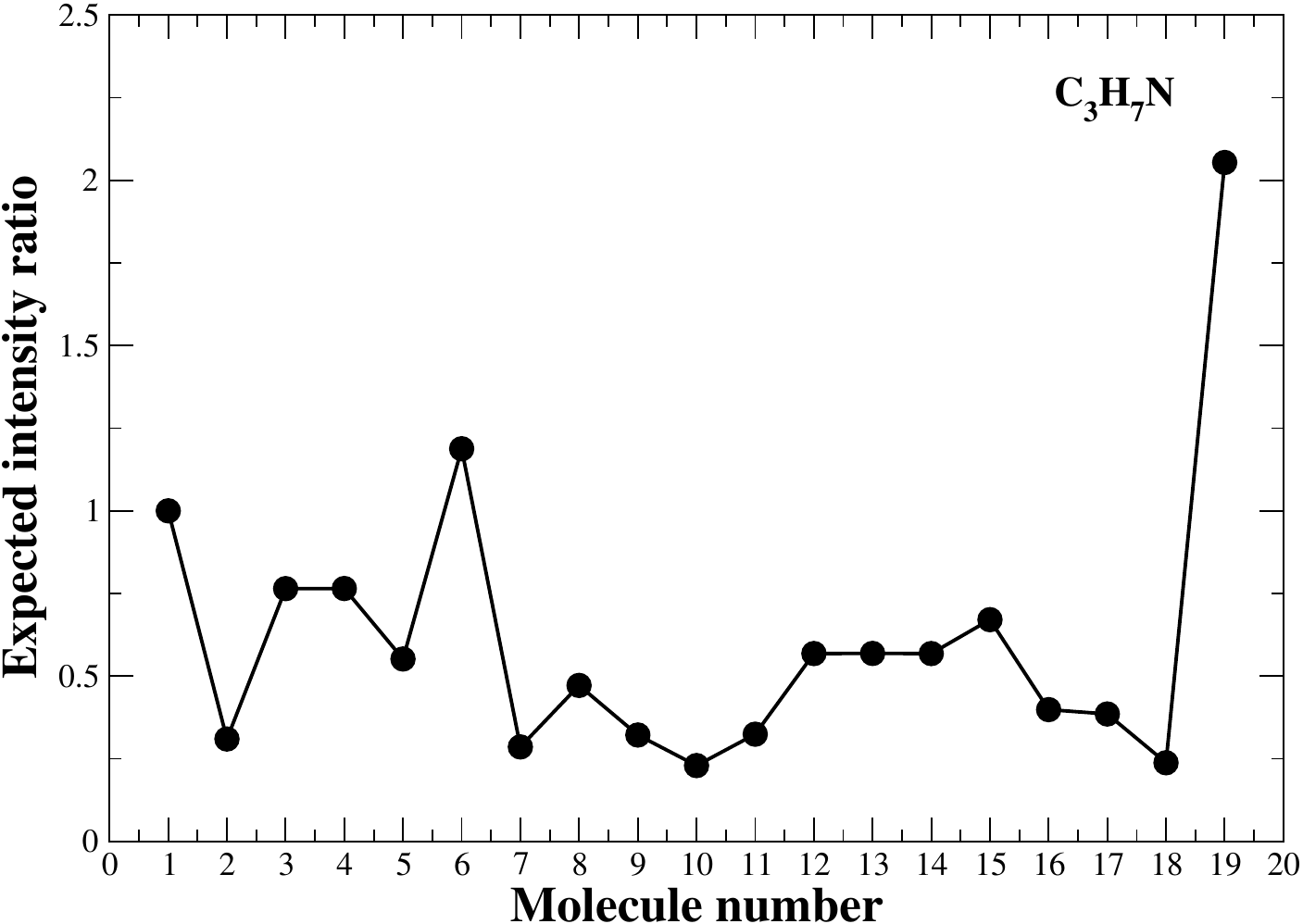}
\caption{Expected intensity ratio of $\rm{C_3H_7N}$ isomeric group by considering the effective dipole moment of all the species.}
\end{figure}

(1Z)-1-propanimine may be formed via two sequential H addition reaction on ice 
with propionitrile ($\rm{CH_3CH_2CN}$), 
where propionitrile may be produced by the radical-radical barrier-less interaction between 
$\mathrm{C_2H_5}$ and CN. Instead of radical reactant CN, $\mathrm{H_2CN}$ may also react 
with $\mathrm{C_2H_5}$ to form propanimine directly by the reaction R19. This reaction is
assumed to be barrier-less in nature.  
We found that the first step (R21) of hydrogen addition with propionitrile has a barrier of
$2712$ K and the second step (R22) is a radical-radical interaction. We assume that the
second step of this sequence is barrier-less. For the gas phase hydrogenation reaction (G21), 
our calculated $\Delta G \ddag$ parameter is found to be  $11.03$ Kcal/mol. 
In Figure 2ab and Figure 3, we have shown the time evolution of 
propanimine (1Z-1-propanimine).
It is evident from the figure 
that the production of (1Z)-1-propanimine is only
favourable in the hot core region. We have a peak gas phase abundance of (1Z)-1-propanimine (propanimine)
$2.20 \times 10^{-08}$. 

\subsection{${C_3H_9N}$ Isomeric Group}

\begin{figure}
\centering
\includegraphics[width=0.45\textwidth]{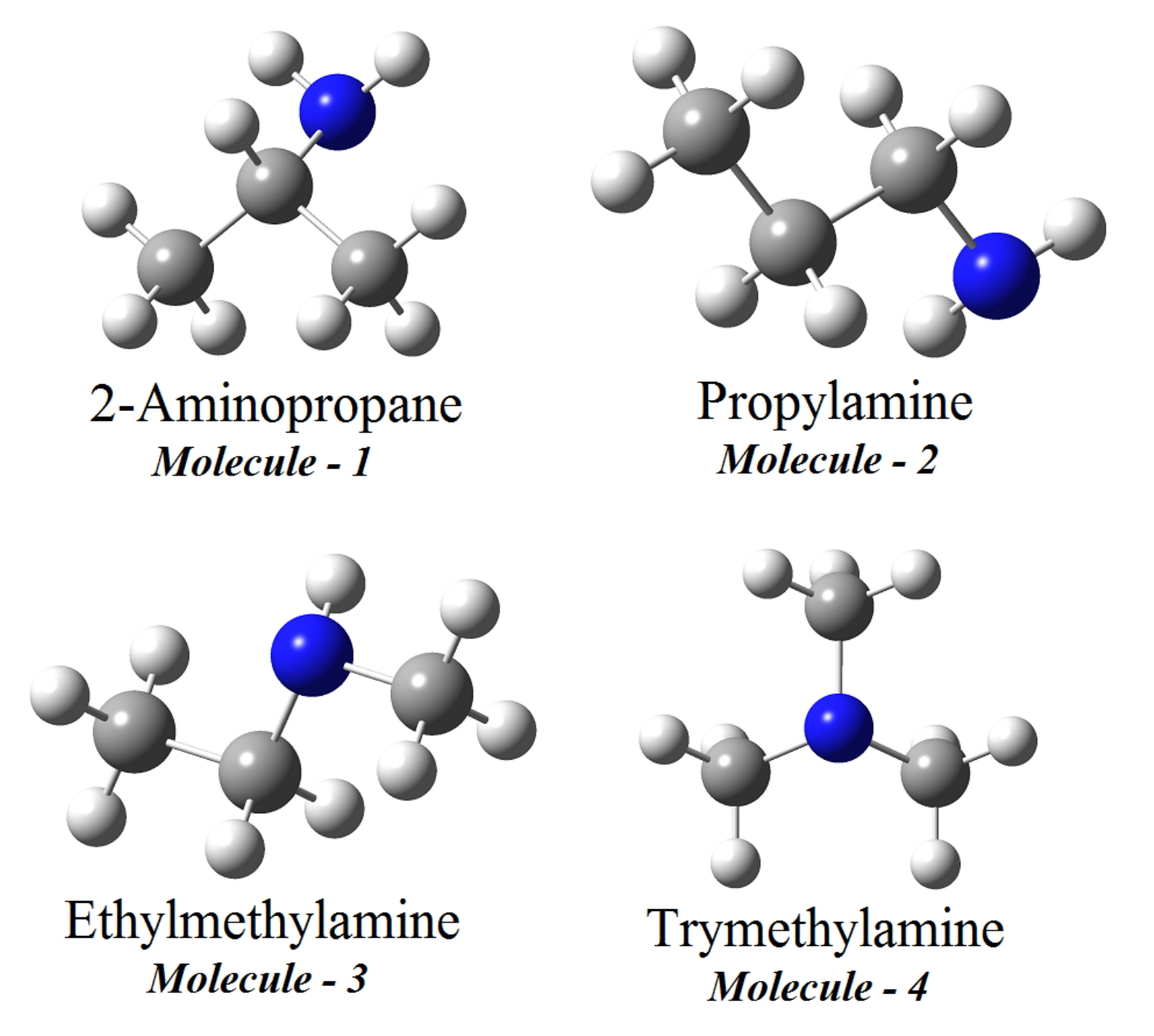}
\caption{$\rm{C_3H_9N}$ isomers.}
\end{figure}

\begin{figure}
\centering
\includegraphics[width=0.45\textwidth]{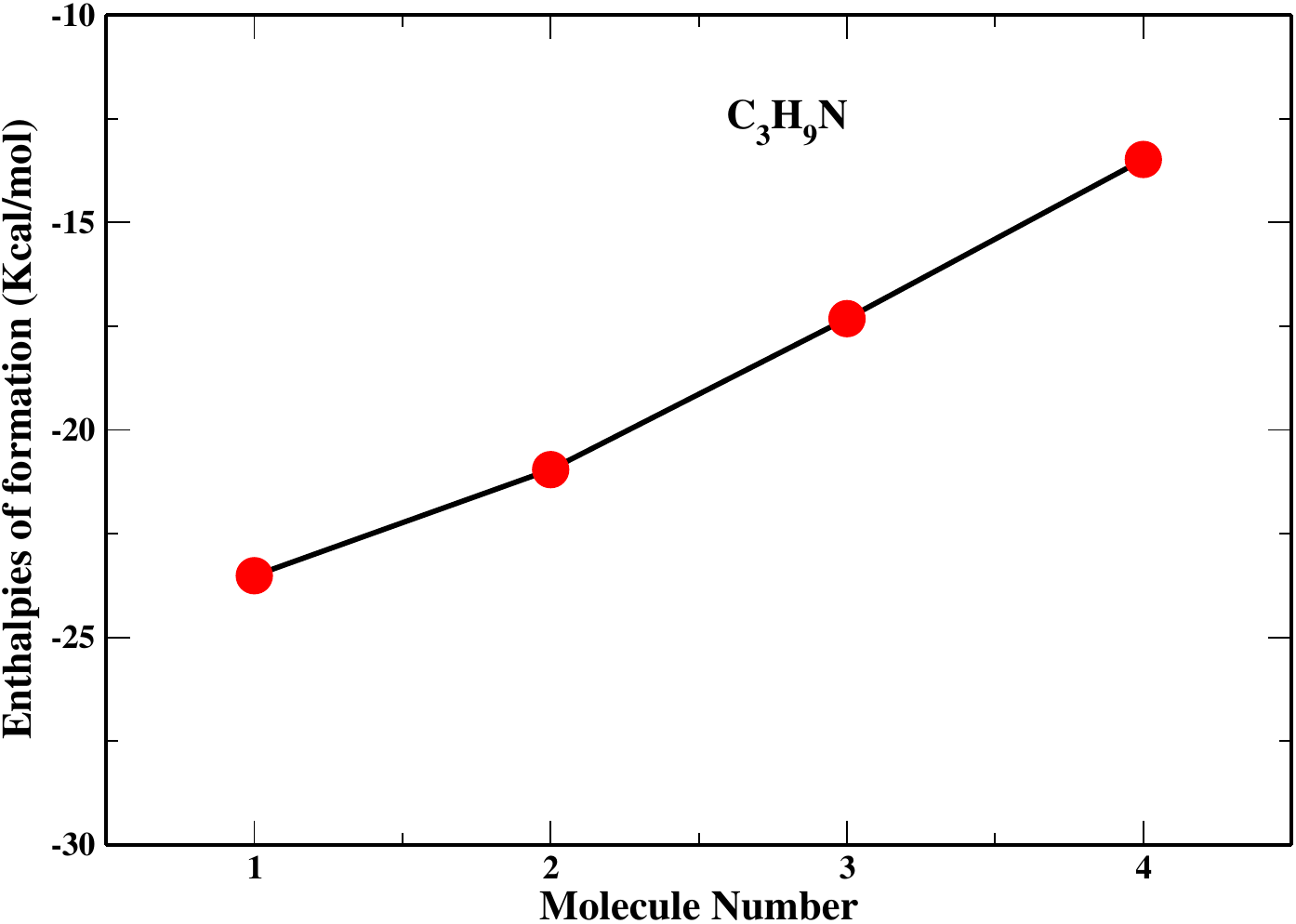}
\caption{Enthalpy of formation of $\rm{C_3H_9N}$ isomeric group.}
\end{figure}

\begin{figure}
\centering
\includegraphics[width=0.45\textwidth]{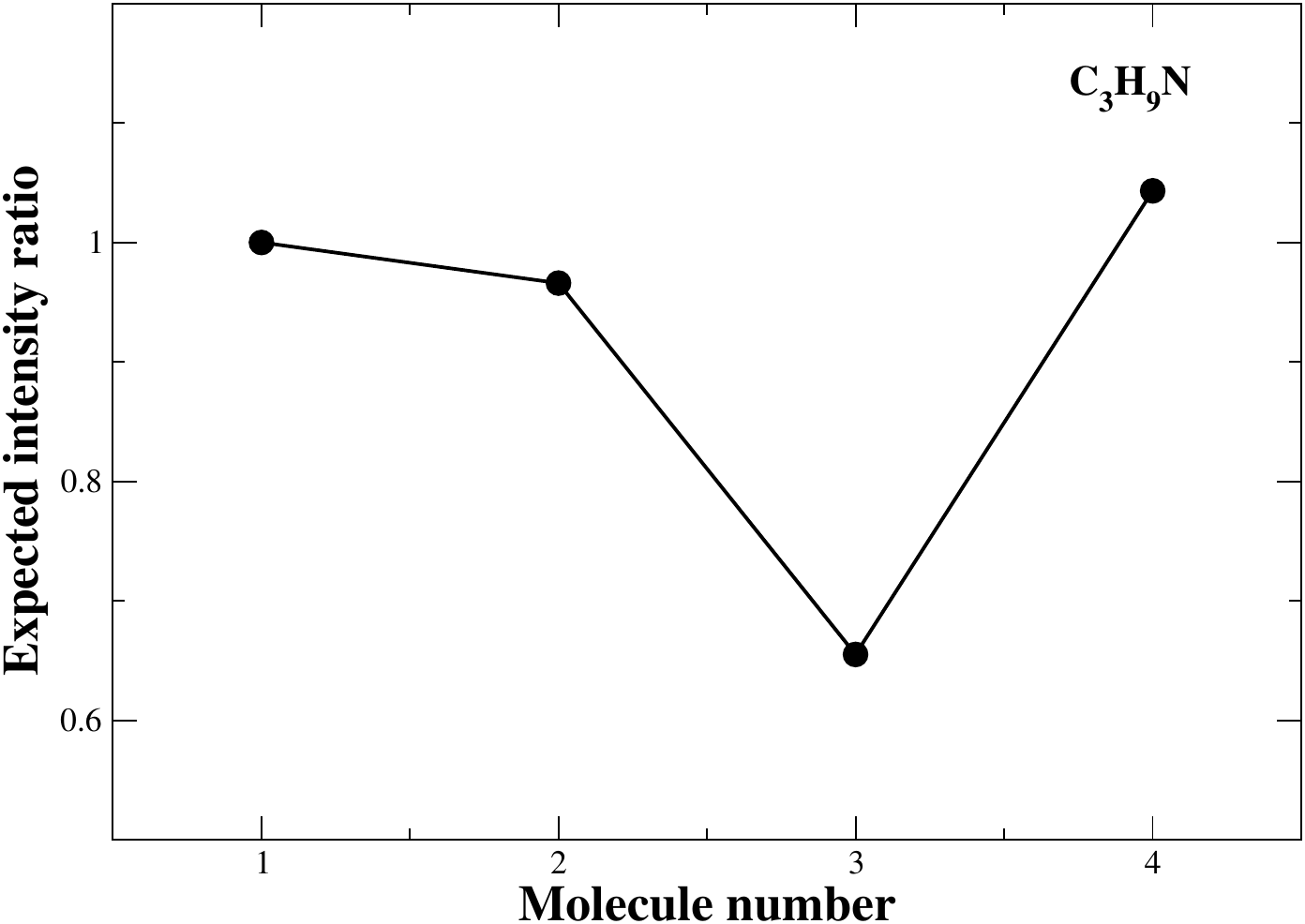}
\caption{Expected intensity ratio of $\rm{C_3H_9N}$ isomeric group
by considering the effective dipole moment of all the species.}
\end{figure}

We consider $4$ species (Figure 14) from this isomeric group namely, 2-aminopropane, propylamine,
ethylmethylamine and trimethylamine. 
Methylamine is the most stable isomer of the $\rm{CH_5N}$ group and ethylamine is 
the most stable isomer of the $\rm{C_2H_7N}$ isomeric group. In general, it is expected that the branched chain molecules 
would be comparatively more stable than the other species of an isomeric group. Recently, \cite{etim17} showed that 
Isopropyl cyanide, a branched chain molecule is the most stable within $\mathrm{C_4H_7N}$ 
isomeric group and Tert-butyl cyanide, another branched chain molecule is the most stable 
species within $\mathrm{C_5H_9N}$ isomeric group. Following the similar trend, we found 
that 2-aminopropane, a branched chain molecule of $\mathrm{C_3H_9N}$ isomeric group 
is the most stable isomer of this group. 2-aminopropane is found to be $2.51$ Kcal/mol 
more stable than the propylamine.
In Figure 15, we have shown the enthalpy of formation of these four species. 
Relative energy and 
enthalpy of formation of these four isomers are shown in Table 2 and arranged based on their 
enthalpy of formation. Theoretically calculated and experimentally obtained enthalpies of formation values
have a similar trend. From Table 2, it is evident that the calculated enthalpies of formation with the
B3LYP/6-31G(d,p) method appears to be comparatively closer to the experimental values than that
of the G4 composite method.
Expected intensity ratio with respect to the species having the minimum enthalpy of formation is shown in 
Figure 16. Interestingly, though trimethylamine which has the lowest total dipole moment value among this
isomeric group, the rotational intensity is found to be maximum  because of its lower partition function. More 
interestingly, due to its unique structure, rotational constants A and B have the same value 
($8.75934$ GHz). We also have reconfirmed this unique nature of trimethylamine (an oblate
symmetric top species) by using the G4 composite method and HF/6-31G(3df) method.
Since the production of propanimine from the very last isomeric group of 
this sequence, $\rm{C_3H_7N}$, is not significantly higher, we have not prepared any reaction 
pathways for the formation of any species from the $\rm{C_3H_9N}$ isomeric group. 

In this section, we have investigated about the chemical abundances of some specific groups of
amine and aldimines. Results presented here, clearly shows that imines (methanimine,
ethanimine and propanimine) and amines (methylamine and ethylamine) may be efficiently
produced in the ice phase (either in the isothermal or in the hot core regime). Depending
upon the barrier energy considered, Figure 3 depicts a nice trend of sublimation. For example,
we have adopted binding energy of ethylamine and methylamine to be $6480$ K and $6584$ K
respectively. Maintaining the trend, ethylamine starts to sublime faster than methylamine. Similarly,
maintaining the trend of binding energies ($5534$ K, $5580$ K and $6337$ K for methanimine, 
ethanimine and propanimine respectively) sublimation sequence is also obtained for the imines as well.

\begin{figure*}
\centering
\includegraphics[width=0.85\textwidth]{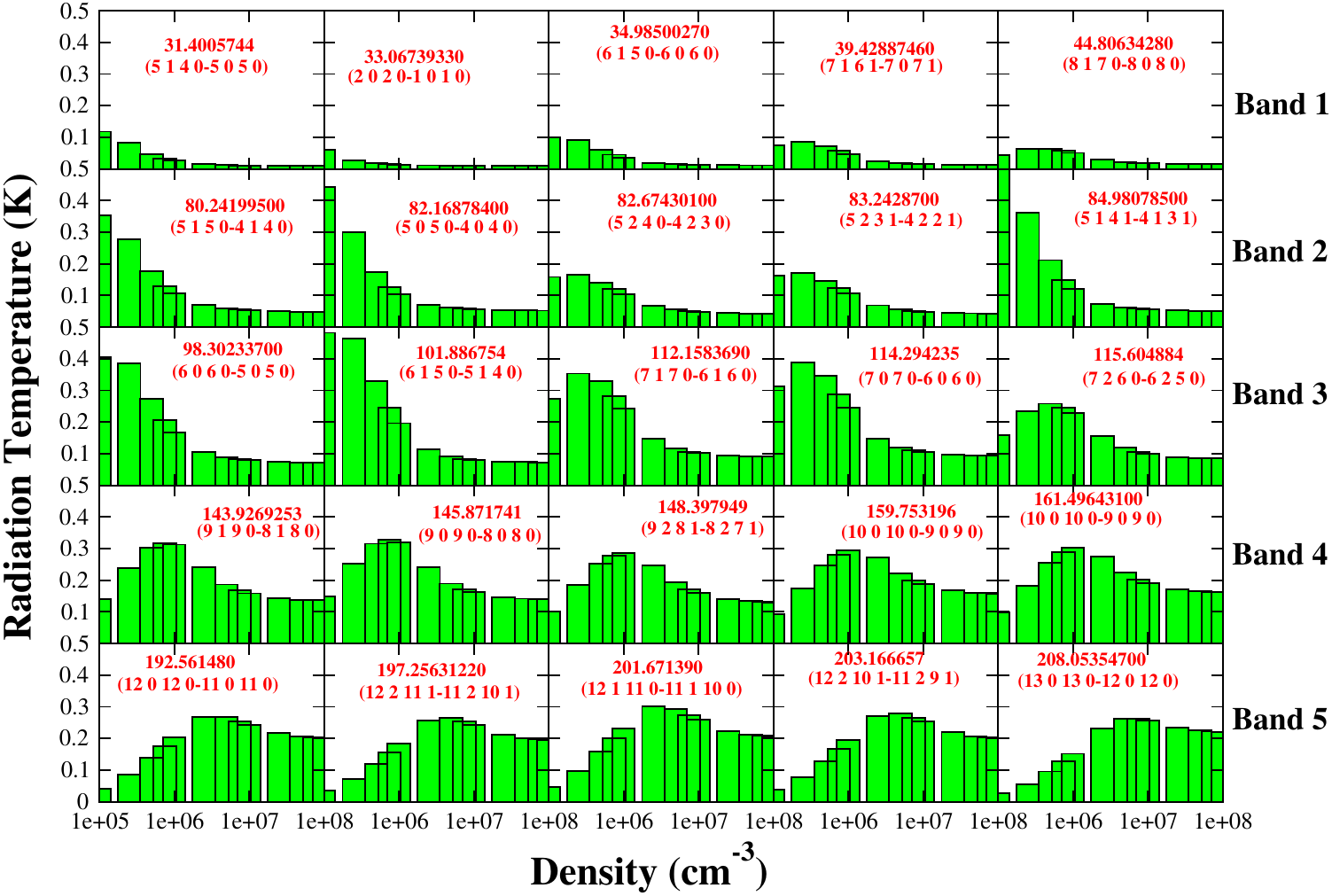}
\caption{Density variation of the intensity of various transitions of ethylamine by considering non-LTE 
approximation.}
\end{figure*}

\begin{figure*}
\centering
\includegraphics[width=0.85\textwidth]{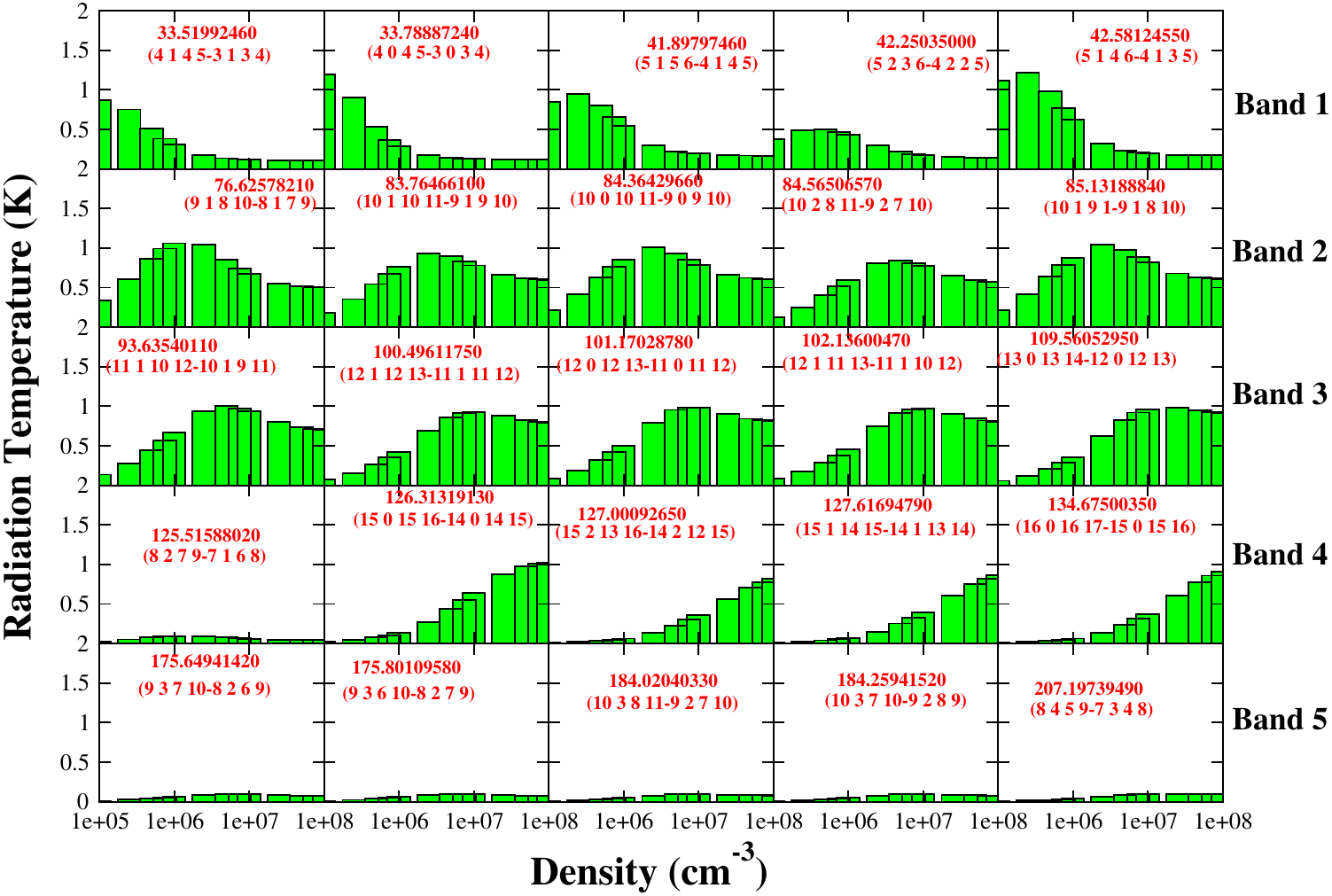}
\caption{Density variation of the intensity of various transitions of propanimine by considering non-LTE 
approximation.}
\end{figure*}

\section{Astrophysical Implications of This Work}
Methanimine is an important prebiotic molecule which is believed to be the precursor
molecule for the formation of simplest amino acid, glycine. It was mentioned earlier that
this species is detected in the ISM. Most interestingly, this species is also detected in the 
upper atmosphere of Titan (the massive moon of Saturn) \citep{vuit06}. 
Present atmosphere of Titan resembles the primeval atmosphere of Earth and thus thought to be
important for the abiotic synthesis. 
Our present study found that methanimine may be further processed to form methylamine which is
yet to be observed in the Titan's atmosphere. 
Modeling the Titan's atmosphere is out of scope for this paper. However, the inclusion of 
proposed pathways in the modeling of Titan atmosphere may come up with the higher mixing ratios of 
higher order imines and amines in Titan's atmosphere. 

We have performed radiative transfer modeling (both with LTE and non-LTE 
consideration) which may be useful for the future astronomical observation of ethylamine and propanimine 
in the ISM. 
For the calculation of the line parameters by using LTE approximation, we use
CASSIS interactive spectrum analyzer (\url{http://cassis.irap.omp.eu/}).
In Table 6, we have pointed out some of the most intense transitions 
of ethylamine that fall in ALMA Bands 1-5.
Required spectroscopic details for ethylamine is available in 
\url{https://www.astro.uni-koeln.de/cdms/catalog}
Similarly, in Table 7, intense transitions of propanimine is shown.
Required spectroscopic details for (1Z)-1-propanimine is obtained
by including the experimentally obtained rotational and distortional constants in
SPCAT program \citep{pick91}.
For preparing these tables, we have used column density of 
H$_2= 10^{23}$ cm$^{-2}$, $\rm{n_H}=10^7$ cm$^{-3}$, 
excitation temperature $= 130$ K, FWHM$= 10$ km/s, $V_{LSR}=64$ km/s, 
source size $=3''$, Abundance of ethylamine =$4.0 \times 10^{-08}$ and abundance of
propanimine to be $2 \times 10^{-8}$. 

We have also performed non-LTE calculation by using RADEX programme \citep{vand07}.
Collisional data file for ethylamine and propanimine is yet to be available in any database. 
Thus, we prepare the collisional data file in the appropriate
format from the spectral information available in JPL (for trans-ethylamine) and from
our calculation (for (1Z)-1-propanimine). 
Altogether we have considered transitions between $251$ energy levels.
Here, we assume H$_2$ as the
colliding partner. 
In order to estimate the line profile with non-LTE, here, we have estimated the collisional
rate of ethylamine and propanimine 
by following the relation mentioned in \cite{shar01}. \cite{shar01} estimated the
collisional rate coefficient for a downward transition of an asymmetric top molecule, cyclopropene at
temperature `T' by,
\begin{equation}
C(J''K_a''K_c'' \rightarrow J' K_a' K_c')= [1 \times 10^{-11}/(2J''+1)]\sqrt{T/30}.
\end{equation}
In Table 8 and Table 9, we have pointed out most intense transitions of trans-ethylamine and 
(1Z)-1-propanimine respectively which are falling within the ALMA band 1-5. 
For the non-LTE calculations, we have used column density of ethylamine to be $10^{15}$ cm$^{-2}$ 
and column density of propanimine to be $5.0 \times 10^{14}$ cm$^{-2}$,
$\rm{n_H}=10^7$ cm$^{-3}$, 
excitation temperature $= 130$ K, FWHM$= 10$ km/s.

For the transitions pointed out in Table 8 and Table 9, we have studied the density variation of 
ethylamine (Figure 17) and propanimine (Figure 18) with the non-LTE consideration. Figures 17 and 18 would be
served as a very useful starting point for  the observation of ethylamine and propanimine in the
ISM. It is to be noted that in absence of the measured or calculated collisional data file, 
we have used our estimated collisional rate but it is known that the non-LTE transitions are heavily dependent
upon collisional rates and consideration of random rates may end up with some misleading results.

\section{Conclusions}
In this work, we examine the possibility of detecting various molecules
that belong to six specific isomeric groups. 
We have used the chemical abundance, enthalpy of formation, optimized 
energy, and expected intensity ratio to shortlist some species that might be the viable candidates
for future astronomical detection in the ISM.
According to our calculation, ethylamine 
remains the most suitable candidate for future astronomical detection in the ISM.
(1Z)-1-propanimine also may be a potential candidate. 
From our gas-grain chemical modeling, we see that the 
precursor molecules (methylamine and ethylamine) of glycine
could efficiently be formed around the star-forming region.
Moreover, radiative transfer modeling (LTE and non-LTE) has been employed 
for the detection of these species in the ISM.

\acknowledgments
{M.S. gratefully acknowledges DST, the Government of India
for providing financial assistance through DST-INSPIRE
Fellowship [IF160109] scheme. A.D. acknowledges the
ISRO respond (grant no. ISRO/RES/2/402/16-17) and DST project (Grant No. SB/S2/HEP-021/2013) for financial support.
B.B. acknowledges DST-INSPIRE Fellowship [IF170046] for providing partial financial assistance.} 

\software{Gaussian 09 \citep{fris13}, SPCAT \citep{pick91}, CASSIS (http://cassis.cesr.fr), 
RADEX \citep{vand07}}


\clearpage

\begin{deluxetable*}{cccccc}
\tablecaption{Enthalpy of Formation and Electronic Energy (E0) with Zero Point Energy (ZPE) 
and Relative Energy (in bracket) with G4 Composite Method
for All Species of Six Isomeric Groups.}
\tablewidth{0pt}
\tabletypesize{\scriptsize} 
\tablehead{
\colhead{\bf Number} & \colhead{\bf Species} & \colhead{\bf Astronomical Status} & \colhead{\bf E0+ZPE} & \colhead{\bf Calculated $\mathrm{\bf \Delta_fH^0}$} & \colhead{\bf Experimental $\mathrm{\bf \Delta_fH^0}$} \\
\colhead{} & \colhead{} & \colhead{} & \colhead{\bf in Hartree/particle} & \colhead{\bf using G4 composite method} &\colhead{\bf (in Kcal/mol)} \\
\colhead{} & \colhead{} &\colhead{}&\colhead{\bf (Relative Energy}&\colhead{\bf (using B3LYP/6-31G(d,p) method)} & \colhead{} \\
& \colhead{} & \colhead{} & \colhead{\bf in Kcal/mol)} & \colhead{\bf (in Kcal/mol)} & \colhead{}
}
\startdata
\multicolumn{6}{c}{$\mathrm{\bf CH_3N}$ {\bf Isomeric Group}} \\
\hline
 1. & Methanimine & observed$^b$ & -94.596377 (0.00)  & 18.2604366 (20.0748878) & --- \\
 2. & $\lambda^1$-Azanylmethane & not observed & -94.519754 (48.08) & 66.3715977 (66.9874996) & --- \\
\hline
\multicolumn{6}{c}{$\mathrm{\bf CH_5N}$ {\bf Isomeric Group}} \\
\hline
 1. & Methylamine & observed$^{c,d}$ & -95.802182 (0.00) & -9.00194363 (-7.3082602) &  -5.37763$^a$ \\
\hline
\multicolumn{6}{c}{$\mathrm{\bf C_2H_5N}$ {\bf Isomeric Group}} \\
\hline
1. & E-Ethanimine & observed$^e$ & -133.896198 (0.00) & 5.90189865 (7.9892830) & 5.74$^f$  \\
2. & Z-Ethanimine & observed$^e$ & -133.895732 (0.29) & 6.20797719 (8.3293932) & ---  \\
3. & Ethenamine & not observed & -133.889919 (3.94) & 9.8284533 (12.7953785) & --- \\
4. & N-Methylmethanimine & not observed & -133.884403 (7.40) & 13.275162 (15.5507728) & 10.51625$^f$  \\
5. & Aziridine & not observed & -133.862508 (21.14) & 26.62315 (29.3566098) &  30.11472$^a$  \\
\hline
\multicolumn{6}{c}{$\mathrm{\bf C_2H_7N}$ {\bf Isomeric Group}} \\
\hline
1. & Ethylamine (trans)  & not observed & -135.094044 (0.00) & -16.366079 (-14.5306661) &  --- \\
2. & Ethylamine (gauche) & not observed & -135.09341 (0.40) & -15.955933 (-14.1008221) & -11.3528$^a$ \\
3. & Dimethylamine & not observed & -135.084612 (5.92) & -10.437412 (-8.4344111) &  -4.445507$^a$ \\
\hline
\multicolumn{6}{c}{$\mathrm{\bf C_3H_7N}$ {\bf Isomeric Group}} \\
\hline
1. & 2-Propanimine & not observed & -173.193699 (0.00) & -4.7991787 (-2.2671314) & ---  \\
2. & 2-Propenamine & not observed & -173.18563 (5.06) & 0.18444024 (3.5818848) & --- \\
3. & (1E)-1-Propanimine & not observed & -173.183877 (6.163) & 1.287921 (3.7186819) & --- \\
4. & (1Z)-1-Propen-1-amine & not observed & -173.183875 (6.164) & 1.2981179 (3.7205644) & --- \\
5. & (1E)-N-Methylethanimine & not observed & -173.183821 (6.20) & 1.4335242 (4.0067087) & --- \\
6. & (1Z)-1-Propanimine & not observed & -173.183423 (6.45) & 1.59211071 (4.0393392) & --- \\
7. & (1E)-1-Propen-1-amine & not observed & -173.181835 (7.44) & 2.6644726 (5.9243778) & --- \\
8. & N-Ethylmethanimine & not observed & -173.17536 (11.51) & 6.5932994 (12.4893824) & ---  \\
9. & N-Methylethenamine & not observed & -173.171735 (13.78) & 9.0000089 (12.4893824) & --- \\
10. & Allylamine & not observed & -173.168277 (15.95) & 11.096984 (14.2338589) & ---  \\
11. & Cyclopropanamine & not observed & -173.159802 (21.27) & 15.985053 (18.7632226) & 18.475$^a$  \\
12. & S-2-Methylaziridine & not observed & -173.157389 (22.7846) & 17.531164 (20.4380455) & --- \\
13. & (2S)-2-Methylaziridine & not observed & -173.157388 (22.7853) & 17.535559 (20.4405556) & --- \\
14. & 2-Methylaziridine (trans) & not observed & -173.157386 (22.7865) & 17.543877 (20.4393006) & --- \\
15. & 2-Methylaziridine (cis) & not observed & -173.156991 (23.03) & 17.7562713 (20.7574479) & --- \\
16. & Azetidine & not observed & -173.1536 (25.16) & 19.607872 (22.3205741) & --- \\
17. & Methylaziridine & not observed & -173.147784 (28.81) & 23.520139 (26.5719511) & --- \\
18. & N-Methylethanamine & not observed & -173.126259 (42.32) & 37.837958 (41.5192279) & --- \\
19. & (Dimethyliminio)methanide & not observed & -173.112784 (50.77) & 45.881231 (50.7837785) & --- \\
\hline
\multicolumn{6}{c}{$\mathrm{\bf C_3H_9N}$ {\bf Isomeric Group}} \\
\hline
1. & 2-Aminopropane & not observed & -174.385779 (0.00) & -23.5149351 (-21.4656727) & -20.0048$^a$ \\
2. & Propylamine & not observed & -174.381773 (2.51) & -20.9566988 (-18.7880896) & -16.7543$^a$  \\
3. & Ethylmethylamine & not observed & -174.375953 (6.16) & -17.3159833 (-15.0857834) & ---  \\
4. & Trimethylamine & not observed & -174.369667 (10.11) & -13.4808824 (-10.9573983) & -5.64054$^a$  \\
\enddata
\tablecomments{
Additional computation of enthalpy of formation by the B3LYP/6-31G(d,p) level of theory is pointed out in parentheses. \\
$^a$ \cite{fren94}, $^b$ \cite{godf73}, $^c$ \cite{kaif74}, $^d$ \cite{four74}, $^e$ \cite{loom13}, \\
$^f$ NIST Chemistry Webbook (\url{http://webbook.nist.gov/chemistry})}
\end{deluxetable*}

\begin{deluxetable*}{cccccc}
\tablecaption{Calculated Dipole Moment Components for All Species of Six Isomeric Groups with HF/6-31G(3df) Method.}
\tablewidth{0pt}
\tabletypesize{\scriptsize} 
\tablehead{
\colhead{\bf Number} & \colhead{\bf Species} & \colhead{\bf $\mu_a$ (in Debye)} & \colhead{\bf $\mu_b$ (in Debye)} & \colhead{\bf $\mu_c$ (in Debye)} & \colhead{\bf $\mu_{tot}$ (in Debye)}
}
\startdata
\multicolumn{6}{c}{$\mathrm{\bf CH_3N}$ {\bf Isomeric Group}} \\
\hline
 1. & Methanimine & $-1.5115 \ [-1.4708] \ (-1.300^g$)  & $-1.4556 \ [-1.4968] \ (-1.500^g$) & 0.0000 [0.0000] (0.000$^g$) & 2.0985 (2.000$^g$)  \\
 2. & $\lambda^1$-Azanylmethane & $-1.9499 \ [-1.9556]$ & $-0.0176 \ [-0.0000]$ & $0.1483 \ [-0.0000]$ & 1.9556 \\
\hline
\multicolumn{6}{c}{$\mathrm{\bf CH_5N}$ {\bf Isomeric Group}} \\
 \hline
 1. & Methylamine & $0.4410 \ [-0.2516]$ & 0.2771 [0.0000] & 1.1774 $[-1.2626]$ & 1.2874 (1.310$^h$) \\
 \hline
\multicolumn{6}{c}{$\mathrm{\bf C_2H_5N}$ {\bf Isomeric Group}} \\
\hline
1. & (E)-Ethanimine & $0.2063 \ [-0.8836]$ & $-2.0884 \ [-1.9256]$ & $0.2912 \ [-0.0000]$ & 2.1187 (1.900$^h$)\\
2. & (Z)-Ethanimine & 0.6062 [2.5103] & $-2.3957 \ [-0.3950]$ & 0.5926 [0.0020] & 2.5412 \\
3. & Ethenamine & 0.5514 [0.9302] & 1.0539 [0.0309] & $-0.7109 \ [-1.0266]$ & 1.3857 \\
4. & N-Methylmethanimine & $-0.2812 \ [-0.1599]$ & $1.0514 \ [-1.6496]$ & $1.2499 \ [-0.0000]$ & 1.6573 (1.530$^i$) \\
5. & Aziridine & 1.6649 [0.0007] & $-0.2522 \ [-0.9880]$ & $0.1768 \ [-1.3749]$ & 1.6931 (1.90$\pm$0.01$^h$) \\
 \hline
\multicolumn{6}{c}{$\mathrm{\bf C_2H_7N}$ {\bf Isomeric Group}} \\
 \hline
1. & Ethylamine (trans) & 0.8802 [0.9708] & $-0.1949 \ [0.8132]$ & $0.8894 \ [-0.0000]$ & 1.2664 (1.304$\pm$0.011$^h$) \\
2. & Ethylamine (gauche) & $-0.5839$ [0.1620] & $-1.0489 \ [-0.6498]$ & $-0.2731 \ [-1.0335]$ & 1.2315 (1.220$^h$)\\
3. & Dimethylamine & $0.1499 \ [-0.0000]$ & $-0.9771 \ [-0.2641]$ & $-0.1444 \ [0.9635]$ & 0.9991 (1.030$^h$) \\
 \hline
\multicolumn{6}{c}{$\mathrm{\bf C_3H_7N}$ {\bf Isomeric Group}} \\
\hline
1. & 2-Propanimine & $1.1894 \ [1.3134]$ & $-1.9000 \ [-2.0677]$ & 0.9879 [$-0.0005$] & 2.4496 \\
2. & 2-Propenamine & $0.4017 \ [-0.1912]$ & 0.4670 [0.9101] & $-1.1975 \ [-0.9739]$ & 1.3467 \\
3. & (1E)-1-Propanimine & $-1.0918 \ [1.0653]$ & $-1.5598 \ [-0.7588]$ & 0.8831 [1.6414] & 2.0987 \\
4. & (1Z)-1-Propen-1-amine & $-0.1209 \ [-1.0650]$ & $1.0920 \ [-0.7572]$ & $1.7882 \ [-1.6423]$ & 2.0987 \\
5. & (1E)-N-Methylethanimine & $-1.3457 \ [0.4136]$ & $-0.2385 \ [-1.5828]$ & 0.8994 [0.0000] & 1.6360 \\
6. & (1Z)-1-Propanimine & $-2.1065 \ [2.4499]$ & 1.5775 [0.9613] & $-0.0300 \ [-0.0188]$ & 2.6318 \\
7. & (1E)-1-Propen-1-amine & $1.0898 \ [-0.6029]$ & $-0.4583 \ [-0.0139]$ & $-0.2748 \ [1.0534]$ & 1.2138 \\
8. & N-Ethylmethanimine & $-1.4418 \ [0.2444]$ & $-0.6595 \ [-0.8449]$ & $-0.0489 \ [-1.3201]$ & 1.5862 \\
9. & N-Methylethenamine & $-1.2679 \ [1.0808]$ & $-0.0583 \ [0.0768]$ & $-0.0886 \ [0.6669]$ & 1.2723 \\
10. & Allylamine & $-0.1645 \ [-0.7685]$ & $-1.1175 \ [-0.8256]$ & $-0.2388 \ [-0.2466]$ & 1.1545 ($\approx$1.2$^h$) \\
11. & Cyclopropanamine & $0.0564 \ [-0.4994]$ & 0.9405 [0.0003] & $-0.8402 \ [1.1307]$ & 1.2361 (1.190$^g$) \\
12. & S-2-Methylaziridine & 1.3228 [0.1520] & $0.6207 \ [-0.7696]$ & $0.7330 \ [-1.4342]$ & 1.6348 \\
13. & (2S)-2-Methylaziridine & $0.5335 \ [-0.1513]$ & 0.4331 [0.7708] & 1.4840 [1.4344] & 1.6354 \\
14. & 2-Methylaziridine (trans) & $0.6744 \ [-0.1519]$ & 0.0897 [0.7692] & $-1.4863 \ [-1.4343]$ & 1.6346 (1.57$\pm$0.03$^h$) \\
15. & 2-Methylaziridine (cis) & 1.2991 [1.2252] & 0.4159 [0.9733] & 1.1382 [0.8410] & 1.7765 (1.77$\pm$0.09$^h$) \\
16. & Azetidine & $0.2802 \ [-0.4148]$ & 0.2805 [0.0000] & $1.1993 \ [-1.1931]$ & 1.2632 \\
17. & Methylaziridine & 0.8549 [0.1236] & 0.1523 [0.0006] & 0.9647 [1.2920] & 1.2980 \\
18. & N-methylethanamine & $-0.5108 \ [-0.5248]$ & $-1.0613 \ [0.9359]$ & $0.4630 \ [-0.1099]$ & 1.0787 \\
19. & (Dimethyliminio)methanide & $-1.4791 \ [-1.1727]$ & $-2.9791 \ [-3.1306]$ & 0.3350 [0.0000] & 3.3430 \\
 \hline
 \multicolumn{6}{c}{$\mathrm{\bf C_3H_9N}$ {\bf Isomeric Group}} \\
 \hline
1. & 2-Aminopropane & $-0.0991 \ [0.0000]$ & $-0.3068 \ [-0.1375]$ & $1.1727 \ [-1.2084]$ & 1.2162 (1.190$^j$) \\
2. & Propylamine & $-0.9663 \ [0.0479]$ & 0.5630 [0.6180] & $0.3766 \ [-1.0045]$ & 1.1804 (1.170$^h$) \\
3. & Ethylmethylamine & 0.1756 [0.0382] & 0.2227 [$-0.3066$] & $-0.8964 \ [-0.8880]$ & 0.9402 \\
4. & Trimethylamine & 0.1174 [0.0000] & $0.1270 \ [-0.0000]$ & $-0.6598 \ [-0.6821]$ & 0.6821 (0.612$^h$) \\
\enddata
\tablecomments{
The dipole moments were calculated both in the input orientation and to the principal axis orientation defined according to the Pickett program (shown in the square bracket). \\
Experimental values are also shown in parentheses. \\
$^g$ \cite{land74}, $^h$ \cite{nels67}, $^i$ \cite{sast64}, $^j$ \cite{mehr77}.}
\end{deluxetable*}

\begin{deluxetable*}{cccccc}
\tablecaption{Calculated Rotational Constants and Rotational Partition Functions at $200$ K for All Species of Six Isomeric Groups (with MP2/6-311++G(d,p) method).}
\tablewidth{0pt}
\tabletypesize{\scriptsize} 
\tablehead{
\colhead{\bf Number} & \colhead{\bf Species} & \colhead{\bf A (in GHz)} & \colhead{\bf B (in GHz)} & \colhead{\bf C (in GHz)} & \colhead{\bf Rotational Partition} \\
\colhead{} & \colhead{} & \colhead{} & \colhead{} & \colhead{} & \colhead{\bf function  at $200$ K}
}
\startdata
\multicolumn{6}{c}{$\mathrm{\bf CH_3N}$ {\bf Isomeric Group}} \\
\hline
 1. & Methanimine &  195.72173 (196.21116$^k$) & 34.45869 (34.64252$^k$) &  29.30013 (29.35238$^k$) & 0.107265(+04) \\
 2. & $\lambda^1$-Azanylmethane & 157.58498 & 27.56460 & 27.56460 & 0.459337(+03) \\
\hline
\multicolumn{6}{c}{$\mathrm{\bf CH_5N}$ {\bf Isomeric Group}} \\
 \hline
 1. & Methylamine & 103.42705 (103.12861$^o$) & 22.75135 (22.62234$^o$) & 21.85872 (21.69598$^o$) & 0.210246(+04) \\
 \hline
\multicolumn{6}{c}{$\mathrm{\bf C_2H_5N}$ {\bf Isomeric Group}} \\
\hline
1. & (E)-Ethanimine & 52.91394 (52.83537$^k$) &  9.76090 (10.07601$^k$) & 8.68503 (8.70427$^k$) & 0.711944(+04) \\
2. & (Z)-Ethanimine & 50.17305 (49.5815$^k$) &  9.76932 (10.15214$^k$)  & 8.61433 (8.644814$^k$)  & 0.733810(+04) \\
3. & Ethenamine & 55.91347 &  9.98960 &  8.55386 & 0.689840(+04) \\
4. & N-Methylmethanimine & 51.70697 (52.52375$^k$)  & 10.71490 (10.66613$^k$) & 9.39306 (9.37719$^k$)  &  0.660982(+04) \\
5. & Aziridine &  22.81302 (22.73612$^p$ ) & 21.19888 (21.19238$^p$) & 13.43101 (13.38307$^p$) & 0.591642(+04) \\
 \hline
\multicolumn{6}{c}{$\mathrm{\bf C_2H_7N}$ {\bf Isomeric Group}} \\
 \hline
1. & Ethylamine (trans) & 31.90275 (31.75833$^m$) & 8.75819 (8.749157$^m$) & 7.82305 (7.798905$^m$) & 0.101989(+05) \\
2. & Ethylamine (gauche) & 32.46287 (32.423470$^n$) & 8.99003 (8.942086$^n$) & 7.86715 (7.825520$^n$) & 0.995128(+04) \\
3. & Dimethylamine & 34.22904 (34.24222$^q$) & 9.38988 (9.33403$^q$)  & 8.26707 (8.21598$^q$)  & 0.925036(+04) \\
 \hline
\multicolumn{6}{c}{$\mathrm{\bf C_3H_7N}$ {\bf Isomeric Group}} \\
\hline
1. & 2-Propanimine & 9.64694 & 8.48897 & 4.78348 & 0.240917(+05) \\
2. & 2-Propenamine & 9.54753 & 8.97855 & 4.79187 & 0.235267(+05) \\
3. & (1E)-1-Propanimine & 23.30832 & 4.33503 & 4.20885 & 0.231221(+05) \\
4. & (1Z)-1-Propen-1-amine & 23.30833 & 4.33547 & 4.20908 & 0.231203(+05) \\
5. & (1E)-N-Methylethanimine & 38.04146 & 4.07938 & 3.86089 & 0.194801(+05) \\
6. & (1Z)-1-Propanimine & 23.19882 (24.1852684$^l$) & 4.28693 (4.2923639$^l$) & 4.17097 (4.1567893$^l$) & 0.234119(+05) \\
7. & (1E)-1-Propen-1-amine & 38.31516 & 3.85080 & 3.59371 &  0.207076(+05) \\
8. & N-Ethylmethanimine & 24.05782 & 4.60272 & 4.48201 & 0.214037(+05) \\
9. & N-Methylethenamine & 32.00012 & 4.30235 & 4.04446 & 0.202070(+05) \\
10. & Allylamine & 23.65103 & 4.23494 & 4.17205 & 0.233259(+05) \\
11. & Cyclopropanamine &  16.28786 (16.26995$^r$) & 6.72692 (6.72300$^r$) &  5.80201 (5.79533$^r$) & 0.189118(+05) \\
12. & S-2-Methylaziridine & 16.91977 & 6.53608 & 5.76664 & 0.188818(+05) \\
13. & (2S)-2-Methylaziridine & 16.91889 & 6.53525 & 5.76592 & 0.188847(+05) \\
14. & 2-Methylaziridine (trans) & 16.92200 & 6.53613 & 5.76677 &  0.188803(+05) \\
15. & 2-Methylaziridine (cis) & 16.68599 & 6.56078 & 5.81794 & 0.188940(+05) \\
16. & Azetidine & 11.54225 & 11.36812 & 6.70581 & 0.160748(+05) \\
17. & Methylaziridine & 16.41594 & 7.25710 & 6.19112 & 0.175575(+05) \\
18. & N-methylethanamine & 36.98588 & 4.14324 & 3.84501 & 0.196438(+05) \\
19. & (Dimethyliminio)methanide & 10.15847 & 9.14934 & 5.12560 & 0.218464(+05) \\
 \hline
\multicolumn{6}{c}{$\mathrm{\bf C_3H_9N}$ {\bf Isomeric Group}} \\
 \hline
1. & 2-Aminopropane & 8.37627 (8.33183$^j$) & 7.99371 (7.97718$^j$) & 4.67889 (4.63719$^j$) & 0.269396(+05) \\
2. & Propylamine & 25.18613 & 3.74012 & 3.49778 & 0.262689(+05) \\
3. & Ethylmethylamine & 26.06033 & 3.92898 & 3.67799 & 0.245712(+05) \\
4. & Trimethylamine & 8.75934 & 8.75934 & 4.99056 & 0.812259(+04) \\
\enddata
\tablecomments{
Experimentally obtained rotational constants are shown in parentheses. \\
$^j$ \cite{mehr77}, $^k$ \cite{pear77}, $^l$ \cite{marg15}, $^m$ \cite{fisc82}, $^n$ \cite{fisc84}, $^o$ \cite{herz66}, $^p$ \cite{bak71}, $^q$ \cite{woll68}, $^r$ \cite{hend69}.}
\end{deluxetable*}

\clearpage

\appendix

Gas phase reaction pathways for the formation/destruction of some important amines and aldimines (Table 5) along with the line parameters for the observations of Trans-ethylamine and  (1Z)-1-Propanimine under LTE (Tables 6 and 7) as well as non-LTE (Tables 8 and 9) approximation.

 \clearpage

\startlongtable
\begin{deluxetable*}{cccccc}
\tablecaption{Gas Phase Formation and Destruction Pathways.}
\tablewidth{0pt}
\tabletypesize{\scriptsize} 
\tablehead{
\colhead{\bf Reaction number (type)} & \colhead{\bf Reactions} & \colhead{\bf $\alpha$} & \colhead{\bf $\beta$} & \colhead{\bf $\gamma$} & \colhead{\bf Rate coefficient@$10$K} \\
\colhead{} & \colhead{} & \colhead{} & \colhead{} & \colhead{} & \colhead{\bf function  at $200$ K}
}
\startdata
\multicolumn{6}{c}{\bf Formation pathways} \\
\hline
G1(RR) & $\mathrm{N+CH_3\rightarrow CH_2NH}$ & $1.00\times10^{-15}$ & -3.0 & 0.0 & $2.70\times10^{-11}$ \\
G2(RR) & $\mathrm{NH+CH_2\rightarrow CH_2NH}$ & $1.00\times10^{-15}$ & -3.0 & 0.0 & $2.70\times10^{-11}$ \\
G3(RR) & $\mathrm{NH_2+CH\rightarrow CH_2NH}$ & $1.00\times10^{-15}$ & -3.0 & 0.0 & $2.70\times10^{-11}$ \\
G4(NR) & $\mathrm{H+HCN\rightarrow H_2CN}$ ($\Delta {G}\ddag=8.37^a$ Kcal/mol) & -& - & - & --- \\
G5(NR) & $\mathrm{H+HCN\rightarrow HCNH}$ ($\Delta {G}\ddag=10.06^a$ Kcal/mol)& - & - & - & ---  \\
G6(RR) & $\mathrm{H+H_2CN\rightarrow CH_2NH}$ & $1.00\times10^{-15}$ & -3.0 & 0.0 & $2.70\times10^{-11}$ \\
G7(RR) & $\mathrm{H+HCNH\rightarrow CH_2NH}$ & $1.00\times10^{-15}$ & -3.0 & 0.0 & $2.70\times10^{-11}$ \\
G8(NR) & $\mathrm{H+CH_2NH\rightarrow CH_3NH}$ ($\Delta {G}\ddag=7.84^a$ Kcal/mol) & - & - & - & --- \\
G9(NR) & $\mathrm{H+CH_2NH\rightarrow CH_2NH_2}$ ($\Delta {G}\ddag=11.64^a$ Kcal/mol) & - & - & - & --- \\
G10(RR) & $\mathrm{H+CH_3NH\rightarrow CH_3NH_2}$ & $1.00\times10^{-15}$ & -3.0 & 0.0 & $2.70\times10^{-11}$ \\
G11(RR) & $\mathrm{H+CH_2NH_2\rightarrow CH_3NH_2}$ & $1.00\times10^{-15}$ & -3.0 & 0.0 & $2.70\times10^{-11}$ \\
G12(RR) & $\mathrm{H+CH_2CN\rightarrow CH_3CN}$ & $1.00\times10^{-15}$ & -3.0 & 0.0 & $2.70\times10^{-11}$ \\
G13(RR) & $\mathrm{CH_3+CN\rightarrow CH_3CN}$ & $1.00\times10^{-15}$ & -3.0 & 0.0 & $2.70\times10^{-11}$ \\
G14(NR) & $\mathrm{H+CH_3CN\rightarrow CH_3CNH}$ ($\Delta {G}\ddag=10.32^a$ Kcal/mol)& - & - & - & ---  \\
G15(RR) & $\mathrm{H+CH_3CNH\rightarrow CH_3CHNH}$ & $1.00\times10^{-15}$ & -3.0 & 0.0 & $2.70\times10^{-11}$ \\
G16(RR) & $\mathrm{CH_3+H_2CN\rightarrow CH_3CHNH}$ & $1.00\times10^{-15}$ & -3.0 & 0.0 & $2.70\times10^{-11}$ \\
G17(NR) & $\mathrm{H+CH_3CHNH\rightarrow CH_3CH_2NH}$ ($\Delta {G}\ddag=9.98^a$ Kcal/mol)& - & -3.0 & 0.0 & ---  \\
G18(RR) & $\mathrm{H+CH_3CH_2NH\rightarrow CH_3CH_2NH_2}$ & $1.00\times10^{-15}$ & -3.0 & 0.0 & $2.70\times10^{-11}$ \\
G19(RR) & $\mathrm{C_2H_5+H_2CN\rightarrow CH_3CH_2CHNH}$ & $1.00\times10^{-15}$ & -3.0 & 0.0 & $2.70\times10^{-11}$ \\
G20(RR) & $\mathrm{C_2H_5+CN\rightarrow CH_3CH_2CN}$ & $1.00\times10^{-15}$ & -3.0 & 0.0 & $2.70\times10^{-11}$ \\
G21(NR) & $\mathrm{H+CH_3CH_2CN\rightarrow CH_3CH_2CNH}$ ($\Delta {G}\ddag=11.03^a$ Kcal/mol)& - & -3.0 & 0.0 & ---  \\
G22(RR) & $\mathrm{H+CH_3CH_2CNH\rightarrow CH_3CH_2CHNH}$  & $1.00\times10^{-15}$ & -3.0 & 0.0 & $2.70\times10^{-11}$ \\
G23(RR) & $\mathrm{C_2H_5+NH\rightarrow CH_3CHNH+H}$  & $2.75\times10^{-12}$ & 0.0 & 0.0 & $2.75\times10^{-12}$ \\
\hline
\multicolumn{6}{c}{\bf Destruction pathways} \\
\hline
G24(IN) & $\mathrm{C^++CH_3CNH\rightarrow C_2H_3^++HNC}$ & $8.10\times10^{-10}$ & -0.5 & 0.0 & $4.44\times10^{-9}$ \\
G25(IN) & $\mathrm{H^++CH_3CNH\rightarrow C2H_4N^++H}$ & $2.50\times10^{-9}$ & -0.5 & 0.0 & $1.37\times10^{-8}$ \\
G26(IN) & $\mathrm{H^++CH_3CNH\rightarrow CH_3CN^++H_2}$  & $2.50\times10^{-9}$ & -0.5 & 0.0 & $1.37\times10^{-8}$ \\
G27(IN) & $\mathrm{He^++CH_3CNH\rightarrow He+HNC^++CH_3}$  & $1.80\times10^{-9}$ & -0.5 & 0.0 & $9.86\times10^{-9}$ \\
G28(IN) & $\mathrm{He^++CH_3CNH\rightarrow He+HNC+CH_3^+}$ & $1.80\times10^{-9}$ & -0.5 & 0.0 & $9.86\times10^{-9}$ \\
G29(IN) & $\mathrm{H_3^++CH_3CNH\rightarrow C_2H_5N^++H_2}$  & $1.50\times10^{-9}$ & -0.5 & 0.0 & $8.22\times10^{-9}$ \\
G30(IN) & $\mathrm{H_3O^++CH_3CNH\rightarrow C_2H_5N^++H_2O}$ & $6.80\times10^{-10}$ & -0.5 & 0.0 & $3.72\times10^{-9}$ \\
G31(IN) & $\mathrm{HCO^++CH_3CNH\rightarrow C_2H_5N^++CO}$ & $6.00\times10^{-10}$ & -0.5 & 0.0 & $3.29\times10^{-9}$ \\
G32(IN) & $\mathrm{HCO_2^++CH_3CNH\rightarrow C_2H_5N^++CO_2}$  & $5.30\times10^{-10}$ & -0.5 & 0.0 & $2.90\times10^{-9}$ \\
G33(IN) & $\mathrm{C^++CH_3CHNH\rightarrow C_2H_4^++HNC}$ & $1.70\times10^{-9}$ & -0.5 & 0.0 & $9.31\times10^{-9}$ \\
G34(IN) & $\mathrm{H^++CH_3CHNH\rightarrow C_2H_4N^++H_2}$  & $5.10\times10^{-9}$ & -0.5 & 0.0 & $2.79\times10^{-8}$ \\
G35(IN) & $\mathrm{H^++CH_3CHNH\rightarrow C_2H_5N^++H}$  & $5.10\times10^{-9}$ & -0.5 & 0.0 & $2.79\times10^{-8}$  \\
G36(IN) & $\mathrm{H_3^++CH_3CHNH\rightarrow C_2H_6N^++H_2}$  & $3.00\times10^{-9}$ & -0.5 & 0.0 &  $1.64\times10^{-8}$  \\
G37(IN) & $\mathrm{H_3O^++CH_3CHNH\rightarrow C_2H_6N^++H_2O}$  & $1.40\times10^{-9}$ & -0.5 & 0.0 &  $7.67\times10^{-9}$ \\
G38(IN) & $\mathrm{HCO^++CH_3CHNH\rightarrow C_2H_6N^++CO}$ & $1.20\times10^{-9}$ & -0.5 & 0.0 &  $6.57\times10^{-9}$  \\
G39(IN) & $\mathrm{HCO_2^++CH_3CHNH\rightarrow C_2H_6N^++CO_2}$ & $1.10\times10^{-9}$ & -0.5 & 0.0 &  $6.02\times10^{-9}$  \\
G40(IN) & $\mathrm{H^++CH_2NH_2\rightarrow NH_2+CH_3^+}$  & $1.00\times10^{-9}$ & 0.0 & 0.0 &  $1.00\times10^{-9}$ \\
G41(IN) & $\mathrm{H^++CH_2NH_2\rightarrow NH_2^++CH_3}$ & $1.00\times10^{-9}$ & 0.0 & 0.0 & $1.00\times10^{-9}$ \\
G42(IN) & $\mathrm{H_3O^++CH_2NH_2\rightarrow CH_2NH_3^++H_2O}$ & $1.00\times10^{-9}$ & 0.0 & 0.0 & $1.00\times10^{-9}$ \\
G43(IN) & $\mathrm{HCO^++CH_2NH_2\rightarrow CH_2NH_3^++CO}$ & $1.00\times10^{-9}$ & 0.0 & 0.0 & $1.00\times10^{-9}$ \\
G44(IN) & $\mathrm{He^++CH_2NH_2\rightarrow NH+CH_3^++He}$  & $1.00\times10^{-9}$ & 0.0 & 0.0 & $1.00\times10^{-9}$ \\
G45(IN) & $\mathrm{H^++CH_3NH\rightarrow NH_2^++CH_3}$ & $1.00\times10^{-9}$ & 0.0 & 0.0 & $1.00\times10^{-9}$ \\
G46(IN) & $\mathrm{H^++CH_3NH\rightarrow NH_2+CH_3^+}$ & $1.00\times10^{-9}$ & 0.0 & 0.0 & $1.00\times10^{-9}$ \\
G47(IN) & $\mathrm{H_3O^++CH_3NH\rightarrow CH_2NH_3^++H_2O}$  & $1.00\times10^{-9}$ & 0.0 & 0.0 & $1.00\times10^{-9}$ \\
G48(IN) & $\mathrm{HCO^++CH_3NH\rightarrow CH_2NH_3^++CO}$  & $1.00\times10^{-9}$ & 0.0 & 0.0 & $1.00\times10^{-9}$ \\
G49(IN) & $\mathrm{He^++CH_3NH\rightarrow NH+CH_3^++He}$ & $1.00\times10^{-9}$ & 0.0 & 0.0 & $1.00\times10^{-9}$ \\
G50(IN) & $\mathrm{H^++CH_3NH_2\rightarrow NH_2+CH_4^+}$ & $1.00\times10^{-9}$ & 0.0 & 0.0 & $1.00\times10^{-9}$ \\
G51(IN) & $\mathrm{H^++CH_3NH_2\rightarrow NH_2^++CH_4}$ & $1.00\times10^{-9}$ & 0.0 & 0.0 & $1.00\times10^{-9}$ \\
G52(IN) & $\mathrm{H_3O^++CH_3NH_2\rightarrow CH_3NH_3^++H_2O}$ & $1.00\times10^{-9}$ & 0.0 & 0.0 & $1.00\times10^{-9}$ \\
G53(IN) & $\mathrm{HCO^++CH_3NH_2\rightarrow CH_3NH_3^++CO}$ & $1.00\times10^{-9}$ & 0.0 & 0.0 & $1.00\times10^{-9}$ \\
G54(IN) & $\mathrm{He^++CH_3NH_2\rightarrow NH_2+CH_3^++He}$ & $1.00\times10^{-9}$ & 0.0 & 0.0 & $1.00\times10^{-9}$ \\
G55(IN) & $\mathrm{C^++CH_3CH_2NH_2\rightarrow C_2H_5^++H_2CN}$ & $1.70\times10^{-9}$ & -0.5 & 0.0 & $9.31\times10^{-9}$ \\
G56(IN) & $\mathrm{H^++CH_3CH_2NH_2\rightarrow CH_3CNH^++H_2+H_2}$ & $5.10\times10^{-9}$ & -0.5 & 0.0 & $2.79\times10^{-8}$ \\
G57(IN) & $\mathrm{H^++CH_3CH_2NH_2\rightarrow C_2H_4N^++H_2+H_2}$ & $5.10\times10^{-9}$ & -0.5 & 0.0 & $2.79\times10^{-8}$ \\
G58(IN) & $\mathrm{H_3^++CH_3CH_2NH_2\rightarrow C_2H_6N^++H_2+H_2}$ & $3.00\times10^{-9}$ & -0.5 & 0.0 & $1.64\times10^{-8}$ \\
G59(IN) & $\mathrm{H_3O^++CH_3CH_2NH_2\rightarrow C_2H_6N^++H_2O+H_2}$ & $1.40\times10^{-9}$ & -0.5 & 0.0 & $7.67\times10^{-9}$ \\
G60(IN) & $\mathrm{HCO^++CH_3CH_2NH_2\rightarrow C_2H_6N^++CO+H_2}$ & $1.20\times10^{-9}$ & -0.5 & 0.0 & $6.57\times10^{-9}$ \\
G61(IN) & $\mathrm{HCO_2^++CH_3CH_2NH_2\rightarrow C_2H_6N^++CO_2+H_2}$  & $1.10\times10^{-9}$ & -0.5 & 0.0 & $6.02\times10^{-9}$ \\
G62(IN) & $\mathrm{C^++CH_3CH_2CN\rightarrow C_2H_5^++C_2N}$  & $1.70\times10^{-9}$ & -0.5 & 0.0 & $9.31\times10^{-9}$ \\
G63(IN) & $\mathrm{H^++CH_3CH_2CN\rightarrow CH_3CNH^++CH_2}$  & $5.10\times10^{-9}$ & -0.5 & 0.0 & $2.79\times10^{-8}$ \\
G64(IN) & $\mathrm{H^++CH_3CH_2CN\rightarrow C_2H_4N^++CH_2}$  & $5.10\times10^{-9}$ & -0.5 & 0.0 & $2.79\times10^{-8}$ \\
G65(IN) & $\mathrm{H_3^++CH_3CH_2CN\rightarrow C_2H_6N^++CH_2}$  & $3.00\times10^{-9}$ & -0.5 & 0.0 & $1.64\times10^{-8}$\\
G66(IN) & $\mathrm{H_3O^++CH_3CH_2CN\rightarrow C_2H_6N^++H_2CO}$ & $1.40\times10^{-9}$ & -0.5 & 0.0 & $7.67\times10^{-9}$ \\
G67(IN) & $\mathrm{HCO^++CH_3CH_2CN\rightarrow C_2H_6N^++C_2O}$  & $1.20\times10^{-9}$ & -0.5 & 0.0 & $6.57\times10^{-9}$\\
G68(IN) & $\mathrm{HCO_2^++CH_3CH_2CN\rightarrow C_2H_5^++CO_2+HCN}$ & $1.10\times10^{-9}$ & -0.5 & 0.0 & $6.02\times10^{-9}$ \\
G69(IN) & $\mathrm{C^++CH_3CH_2CNH\rightarrow C_2H_5^++C_2N+H}$  & $1.70\times10^{-9}$ & -0.5 & 0.0 & $9.31\times10^{-9}$ \\
G70(IN) & $\mathrm{H^++CH_3CH_2CNH\rightarrow CH_3CNH^++CH_2+H}$  & $5.10\times10^{-9}$ & -0.5 & 0.0 & $2.79\times10^{-8}$ \\
G71(IN) & $\mathrm{H^++CH_3CH_2CNH\rightarrow C_2H_4N^++CH_3}$  & $5.10\times10^{-9}$ & -0.5 & 0.0 & $2.79\times10^{-8}$ \\
G72(IN) & $\mathrm{H_3^++CH_3CH_2CNH\rightarrow C_2H_6N^++CH_3}$  & $3.00\times10^{-9}$ & -0.5 & 0.0 & $1.64\times10^{-8}$ \\
G73(IN) & $\mathrm{H_3O^++CH_3CH_2CNH\rightarrow C_2H_6N^++H_2CO+H}$ & $1.40\times10^{-9}$ & -0.5 & 0.0 & $7.67\times10^{-9}$ \\
G74(IN) & $\mathrm{HCO^++CH_3CH_2CNH\rightarrow C_2H_6N^++C_2O+H}$  & $1.20\times10^{-9}$ & -0.5 & 0.0 & $6.57\times10^{-9}$ \\
G75(IN) & $\mathrm{HCO_2^++CH_3CH_2CNH\rightarrow C_2H_6N^++C_2O+OH}$  & $1.10\times10^{-9}$ & -0.5 & 0.0 & $6.02\times10^{-9}$ \\
G76(IN) & $\mathrm{C^++CH_3CH_2CHNH\rightarrow C_2H_5^++C_2N+H_2}$ & $1.70\times10^{-9}$ & -0.5 & 0.0 & $9.31\times10^{-9}$ \\
G77(IN) & $\mathrm{H^++CH_3CH_2CHNH\rightarrow CH_3CNH^++CH_3+H}$  & $5.10\times10^{-9}$ & -0.5 & 0.0 & $2.79\times10^{-8}$ \\
G78(IN) & $\mathrm{H^++CH_3CH_2CHNH\rightarrow C_2H_4N^++CH_3+H}$ & $5.10\times10^{-9}$ & -0.5 & 0.0 & $2.79\times10^{-8}$ \\
G79(IN) & $\mathrm{H_3^++CH_3CH_2CHNH\rightarrow C_2H_6N^++CH_3+H}$ & $3.00\times10^{-9}$ & -0.5 & 0.0 & $1.64\times10^{-8}$ \\
G80(IN) & $\mathrm{H_3O^++CH_3CH_2CHNH\rightarrow C_2H_6N^++H_2CO+H_2}$  & $1.40\times10^{-9}$ & -0.5 & 0.0 & $7.67\times10^{-9}$ \\
G81(IN) & $\mathrm{HCO^++CH_3CH_2CHNH\rightarrow C_2H_6N^++C_2O+H_2}$ & $1.20\times10^{-9}$ & -0.5 & 0.0 & $6.57\times10^{-9}$ \\
G82(IN) & $\mathrm{HCO_2^++CH_3CH_2CHNH\rightarrow C_2H_6N^++C_2O+H_2O}$ & $1.10\times10^{-9}$ & -0.5 & 0.0 & $6.02\times10^{-9}$ \\
G83(IN) & $\mathrm{C^++CH_3CH_2NH\rightarrow C_2H_5^++HCN}$ & $1.70\times10^{-9}$ & -0.5 & 0.0 & $9.31\times10^{-9}$ \\
G84(IN) & $\mathrm{H^++CH_3CH_2NH\rightarrow CH_3CNH^++H_2+H}$ & $5.10\times10^{-9}$ & -0.5 & 0.0 & $2.79\times10^{-8}$ \\
G85(IN) & $\mathrm{H^++CH_3CH_2NH\rightarrow C_2H_4N^++H_2+H}$ & $5.10\times10^{-9}$ & -0.5 & 0.0 & $2.79\times10^{-8}$ \\
G86(IN) & $\mathrm{H_3^++CH_3CH_2NH\rightarrow C_2H_6N^++H_2+H}$  & $3.00\times10^{-9}$ & -0.5 & 0.0 &  $1.64\times10^{-8}$ \\
G87(IN) & $\mathrm{H_3O^++CH_3CH_2NH\rightarrow C_2H_6N^++H_2O+H}$ & $1.40\times10^{-9}$ & -0.5 & 0.0 & $7.67\times10^{-9}$ \\
G88(IN) & $\mathrm{HCO^++CH_3CH_2NH\rightarrow C_2H_6N^++CO+H}$ & $1.20\times10^{-9}$ & -0.5 & 0.0 & $6.57\times10^{-9}$ \\
G89(IN) & $\mathrm{HCO_2^++CH_3CH_2NH\rightarrow C_2H_6N^++CO_2+H}$ & $1.10\times10^{-9}$ & -0.5 & 0.0 &  $6.02\times10^{-9}$ \\
G90(NN) & $\mathrm{CH_3CNH+H\rightarrow CH_3CN+H_2}$ & $1.28\times10^{-11}$ & 0.5 & 0.0 & $2.37\times10^{-12}$ \\
G91(NN) & $\mathrm{CH_3CHNH+H\rightarrow CH_3CNH+H_2}$ & $1.28\times10^{-11}$ & 0.5 & 1050.0 & $2.37\times10^{-12}$ \\
G92(NN) & $\mathrm{CH_3CNH+C\rightarrow CH_3CN+CH}$ & $4.18\times10^{-12}$ & 0.5 & 0.0 & $7.67\times10^{-13}$ \\
G93(NN) & $\mathrm{CH_3CHNH+C\rightarrow CH_3CNH+CH}$  & $4.18\times10^{-12}$ & 0.5 & 0.0 & $7.67\times10^{-13}$ \\
G94(NN) & $\mathrm{C_2H_5+N\rightarrow CH_3CNH+H}$  & $8.30\times10^{-12}$ & 0.0 & 0.0 & $8.30\times10^{-12}$ \\
G95(DR) & $\mathrm{C_2H_4N^++e^-\rightarrow CH_3CN+H}$ & $1.50\times10^{-7}$ & -0.5 & 0.0 & $8.22\times10^{-7}$ \\
G96(DR) & $\mathrm{C_2H_4N^++e^-\rightarrow CH_2CN+H+H}$  & $1.50\times10^{-7}$ & -0.5 & 0.0 & $8.22\times10^{-7}$ \\
G97(DR) & $\mathrm{C_2H_5N^++e^-\rightarrow CH_3CNH+H}$ & $1.50\times10^{-7}$ & -0.5 & 0.0 & $8.22\times10^{-7}$ \\
G98(DR) & $\mathrm{C_2H_5N^++e^-\rightarrow CH_3+H_2CN}$  & $1.50\times10^{-7}$ & -0.5 & 0.0 & $8.22\times10^{-7}$ \\
G99(DR) & $\mathrm{C_2H_6N^++e^-\rightarrow CH_3CHNH+H}$  & $1.50\times10^{-7}$ & -0.5 & 0.0 & $8.22\times10^{-7}$ \\
G100(DR) & $\mathrm{C_2H_6N^++e^-\rightarrow CH_4+H_2CN}$ & $1.50\times10^{-7}$ & -0.5 & 0.0 & $8.22\times10^{-7}$ \\
G101(DR) & $\mathrm{CH_2NH_3^++e^-\rightarrow CH_4+NH}$  & $1.50\times10^{-7}$ & -0.5 & 0.0 & $8.22\times10^{-7}$ \\
G102(DR) & $\mathrm{CH_3NH_3^++e^-\rightarrow CH_4+NH_2}$ & $1.50\times10^{-7}$ & -0.5 & 0.0 & $8.22\times10^{-7}$ \\
G103(PH) & $\mathrm{CH_3CNH+PHOTON\rightarrow CH_3+HNC}$ & $1.00\times10^{-9}$ & 0.0 & 1.9 & $5.60\times10^{-18}$ \\
G104(PH) & $\mathrm{CH_3CNH+PHOTON\rightarrow CH_3CN+H}$ & $1.00\times10^{-9}$ & 0.0 & 1.9 & $5.60\times10^{-18}$ \\
G105(PH) & $\mathrm{CH_3CHNH+PHOTON\rightarrow CH_3+H_2CN}$  & $1.00\times10^{-9}$ & 0.0 & 1.9 & $5.60\times10^{-18}$ \\
G106(PH) & $\mathrm{CH_3CHNH+PHOTON\rightarrow CH_3CNH+H}$ & $1.00\times10^{-9}$ & 0.0 & 1.9 & $5.60\times10^{-18}$ \\
G107(PH) & $\mathrm{CH_3CH_2CN+PHOTON\rightarrow CH_3CNH+CH}$ & $1.00\times10^{-9}$ & 0.0 & 1.9 & $5.60\times10^{-18}$ \\
G108(PH) & $\mathrm{CH_3CH_2NH_2+PHOTON\rightarrow CH_3NH_2+CH_2}$ & $1.00\times10^{-9}$ & 0.0 & 1.9 & $5.60\times10^{-18}$ \\
G109(PH) & $\mathrm{CH_3CH_2CNH+PHOTON\rightarrow CH_3CNH+CH_2}$  & $1.00\times10^{-9}$ & 0.0 & 1.9 & $5.60\times10^{-18}$ \\
G110(PH) & $\mathrm{CH_3CH_2CHNH+PHOTON\rightarrow CH_3CNH+CH_3}$ & $1.00\times10^{-9}$ & 0.0 & 1.9 & $5.60\times10^{-18}$ \\
G111(PH) & $\mathrm{CH_2NH_2+PHOTON\rightarrow H_2CN+H_2}$  & $1.00\times10^{-9}$ & 0.0 & 1.6 & $3.94\times10^{-16}$ \\
G112(PH) & $\mathrm{CH_3NH+PHOTON\rightarrow H_2CN+H_2}$ & $1.00\times10^{-9}$ & 0.0 & 1.6 & $3.94\times10^{-16}$ \\
G113(PH) & $\mathrm{CH_3NH_2+PHOTON\rightarrow H_2CN+H_2+H}$  & $3.50\times10^{-9}$ & 0.0 & 1.6 & $3.94\times10^{-16}$ \\
G114(PH) & $\mathrm{CH_3CH_2NH+PHOTON\rightarrow CH_3CHNH+H}$ & $3.50\times10^{-9}$ & 0.0 & 1.6 & $3.94\times10^{-16}$ \\
G115(PH) & $\mathrm{CH_3CH_2NH+PHOTON\rightarrow C_2H_5+NH}$  & $3.50\times10^{-9}$ & 0.0 & 1.6 & $3.94\times10^{-16}$\\
G116(CR) & $\mathrm{CH_3CNH+CRPHOT\rightarrow CH_3+HNC}$ & $1.30\times10^{-17}$ & 0.0 & 1.9 & $1.95\times10^{-14}$ \\
G117(CR) & $\mathrm{CH_3CNH+CRPHOT\rightarrow CH_3CN+H}$ & $1.30\times10^{-17}$ & 0.0 & 1.9 & $1.95\times10^{-14}$ \\
G118(CR) & $\mathrm{CH_3CHNH+CRPHOT\rightarrow CH_3+H_2CN}$  & $1.30\times10^{-17}$ & 0.0 & 1.9 & $1.95\times10^{-14}$ \\
G119(CR) & $\mathrm{CH_3CHNH+CRPHOT\rightarrow CH_3CNH+H}$ & $1.30\times10^{-17}$ & 0.0 & 1.9 & $1.95\times10^{-14}$ \\
G120(CR) & $\mathrm{CH_3CH_2NH_2+CRPHOT\rightarrow CH_3NH_2+CH_2}$ & $1.30\times10^{-17}$ & 0.0 & 1.9 & $1.95\times10^{-14}$ \\
G121(CR) & $\mathrm{CH_3CH_2CN+CRPHOT\rightarrow CH_3CNH+CH}$  & $1.30\times10^{-17}$ & 0.0 & 1.9 & $1.95\times10^{-14}$ \\
G122(CR) & $\mathrm{CH_3CH_2CNH+CRPHOT\rightarrow CH_3CNH+CH_2}$ & $1.30\times10^{-17}$ & 0.0 & 1.9 & $1.95\times10^{-14}$ \\
G123(CR) & $\mathrm{CH_3CH_2CHNH+CRPHOT\rightarrow CH_3CNH+CH_3}$  & $1.30\times10^{-17}$ & 0.0 & 1.9 & $1.95\times10^{-14}$ \\
G124(CR) & $\mathrm{CH_2NH_2+CRPHOT\rightarrow NH+CH_3}$  & $1.30\times10^{-17}$ & 0.0 & 500.0 & $1.95\times10^{-14}$ \\
G125(CR) & $\mathrm{CH_3NH+CRPHOT\rightarrow NH+CH_3}$  & $1.30\times10^{-17}$ & 0.0 & 500.0 & $1.95\times10^{-14}$ \\
G126(CR) & $\mathrm{CH_3NH_2+CRPHOT\rightarrow NH_2+CH_3}$ & $1.30\times10^{-17}$ & 0.0 & 500.0 & $1.95\times10^{-14}$ \\
G127(CR) & $\mathrm{CH_3CH2NH+CRPHOT\rightarrow CH_3CHNH+H}$ & $1.30\times10^{-17}$ & 0.0 & 1.9 & $1.95\times10^{-14}$ \\
G128(CR) & $\mathrm{CH_3CH2NH+CRPHOT\rightarrow C_2H_5+NH}$ & $1.30\times10^{-17}$ & 0.0 & 1.9 & $1.95\times10^{-14}$ \\
\enddata
\tablecomments{
$^a$ This work.}
\end{deluxetable*}

\begin{deluxetable*}{cccc}
\tablecaption{Line Parameters of Trans-ethylamine in the Millimeter and Submillimeter Regime Using ALMA (LTE).}
\tablewidth{0pt}
\tabletypesize{\scriptsize} 
\tablehead{
\colhead{\bf Frequency range (GHz)} & \colhead{\bf Frequency (GHz)$^a$} & \colhead{\bf Transition (J$^\prime$ k$_a^\prime$ k$_c^\prime$ v$^\prime$ - J$^{\prime\prime}$ k$_a^{\prime\prime}$ k$_c^{\prime\prime}$ v$^{\prime\prime}$)} & \colhead{\bf Intensity (K)}
}
\startdata
    & 31.4005744 & 5 1 4 0 - 5 0 5 0& 0.0058\\
    & 33.0673762 & 2 0 2 1 - 1 0 1 1&0.0063\\
31-45 (ALMA Band 1)  &34.9850027 & 6 1 5 0 - 6 0 6 0& 0.008\\
    & 39.4293083 & 7 1 6 0 - 7 0 7 0&0.011 \\
    & 44.8063428 & 8 1 7 0 - 8 0 8 0& 0.015\\
   \hline
    & 80.241995 & 5 1 5 1 - 4 1 4 1& 0.098\\
    &82.168784 & 5 0 5 1 - 4 0 4 1 & 0.106\\
67-90 (ALMA Band 2)  & 82.674301 & 5 2 4 1 - 4 2 3 1&0.088\\
and    & 83.24287 & 5 2 3 1 - 4 2 2 1 &0.089\\
84-116 (ALMA Band 3)       & 84.980785 &5 1 4 1 - 4 1 3 1& 0.107 \\
     & 101.886754 & 6 1 5 1 - 5 1 4 1& 0.169\\
     &112.158369 &7 1 7 1 - 6 1 6 1& 0.222\\
     &114.294235 & 7 0 7 1 - 6 0 6 1 & 0.234\\
     & 115.604884 & 7 2 6 1 - 6 2 5 1 & 0.211\\
     & 115.9771874 & 7 4 3 1 - 6 4 2 1 & 0.256\\
   \hline   
    & 145.871741 & 9 0 9 1 - 8 0 8 1 & 0.38 \\
    &149.11269 & 9 5 5 1 - 8 5 4 1 (9 5 4 1 - 8 5 3 1) & 0.436\\
125-163 (ALMA Band 4)      &152.224777 &9 1 8 1 - 8 1 7 1   & 0.391\\
    & 159.753196 & 10 1 10 1 - 9 1 9 1 & 0.445\\
    & 161.496431 & 10 0 10 1 - 9 0 9 1 & 0.45 \\
   \hline
    &182.3234429 &11 5 7 0 - 10 5 6 0 & 0.684\\
    & 198.94447 &12 5 8 1 - 11 5 7 1   &0.767\\
163-211 (ALMA Band 5) &198.945801 &12 5 7 1 - 11 5 6 1&0.768\\
    &207.02715 & 13 1 13 1 - 12 1 12 1  & 0.64 \\
    & 208.053547 & 13 0 13 1 - 12 0 12 1 & 0.65\\
   \enddata
  \tablecomments{ 
$^a$ For the transitions with same J$^\prime$ k$_a^\prime$ k$_c^\prime$ - $J ^{\prime\prime}$ k$_a^{\prime\prime}$ k$_c^{\prime\prime}$ but having different
vibrational state please see the cat file available in \url{https://www.astro.uni-koeln.de/cdms/catalog}.}
\end{deluxetable*}

\begin{deluxetable*}{cccc}
\tablecaption{Line Parameters of (1Z)-1-Propanimine in the Millimeter and Submillimeter Regime Using ALMA (LTE).}
\tablewidth{0pt}
\tabletypesize{\scriptsize} 
\tablehead{
\colhead{\bf Frequency range (GHz)} & \colhead{\bf Frequency (GHz)$^a$} & \colhead{\bf Transition (J$^\prime$ k$_a^\prime$ k$_c^\prime$ v$^\prime$ - J$^{\prime\prime}$ k$_a^{\prime\prime}$ k$_c^{\prime\prime}$ v$^{\prime\prime}$)} & \colhead{\bf Intensity (K)}
}
\startdata
&33.5198821&4 1 4 3 - 3 1 3 2&0.117\\
&33.7888724&4 0 4 5- 3 0 3 4&0.136\\
&33.7897526&4 2 3 4 - 3 2 2 3&0.105\\
&34.0663468&4 1 3 4 - 3 1 2 3&0.123\\
  31-45 (ALMA Band 1) &41.897946&5 1 5 6 - 4 1 4 5&0.261\\
&42.2302861&5 0 5 6 - 4 0 4 5&0.290\\
&42.2318919& 5 3 3 5 - 4 3 1 5&0.326\\
&42.2361126&5 2 4 6 - 4 2 3 5&0.227\\
&42.5812455& 5 1 4 6 - 4 1 3 5&0.270\\
\hline
&67.0236193&8 1 8 8 - 7 1 7 7&1.110\\
&67.5282969&8 0 8 8 - 7 0 7 7 &2.252\\
&67.5717125& 8 3 6 7 - 7 3 5 8&1.725\\
&68.117366&8 1 7 8 - 7 1 6 7 &1.145\\
&75.3952267&9 1 9 9 - 8 1 8 8&1.510\\
&75.9707109&9 5 4 8 - 8 5 3 7&1.720\\
&75.9497207&9 0 9 10 - 8 2 6 9&1.440\\
   67-90 (ALMA Band 2)&76.0013965&9 4 6 10 - 8 4 49&2.252\\
and &76.0187663&9 3 6 9 - 8 3 5 9&2.244\\
84-116 (ALMA Band 3)   &76.6257696&9 1 8 9 - 8 1 7 8&1.555\\
&83.7646502&10 11 0 10 - 9 1 9 9&1.955\\
&84.364287&10 0 10 9 - 9 1 8 9&2.000\\
&84.4134387& 10 5 6 11 - 9 5 4 10&2.40\\
&84.4483753&10 4 6 9 - 9 4 5 10&4.650\\
&84.4661717&10 3 7 10 - 9 3 6 10&2.450\\
&84.5650434&10 2 8 10 - 9 2 7 9&1.870\\
&85.1318785&10 0 10 9 - 9 1 8 9&2.080\\
&92.1316784&11 1 11 11 - 10 1 10 10&2.420\\
&92.7713594& 11 0 11 12 - 10 0 10 11&2.480\\
&92.8567065&11 5 6 10 - 10 5 6 9&3.150\\
&92.8959487&11 4 8 10 - 10 4 6 9&3.720\\
&93.0411029&11 2 9 11 - 10 2 8 10&3.720\\
&93.6354011&11 1 10 12 - 10 1 9 11&2.480\\
&100.4962033&12 1 12 11 - 11 1 11 12&2.900\\
&101.1702803&12 0 12 12 - 11 0 11 11&2.977\\
&101.3005235&12 5 8 13 - 11 5 6 12&3.925\\
&101.3444369&12 4 8 12 - 12 4 7 11&4.550\\
&101.5225577&12 2 10 11 - 12 2 9 12&2.800\\
&102.1359978&12 1 11 12 - 11 1 10 11&2.960\\
&109.7449013&13 5 9 13 - 12 5 7 12&4.703\\
&109.7938556&13 4 9 13 - 12 4 8 12&5.395\\
&110.0094549&13 2 11 13 - 12 2 10 12&3.300\\
&110.6333677&13 1 12 14 - 12 1 11 13&3.455\\
\hline
&126.6358855&15 5 10 15 - 14 5 10 14&6.200\\
&126.6960852&15 4 11 14 - 14 4 11 13&10.080\\
&133.9242154&16 1 16 15 - 15 0 15 16&4.730\\
&134.6750035&16 0 16 17 - 15 0 15 16&4.800\\
&135.148249&16 4 13 15 - 15 4 12 16&7.750\\
  125-163 (ALMA Band 4)  &143.6023263&17 4 13 17 -16 4 12 16&8.350\\
&151.9783682& 18 5 14 18 - 17 5 12 17&8.100\\
&152.0568371&18 4 15 17 - 17 4 14 18&8.750\\
&160.4277999&19 5 14 18 - 18 5 13 19&8.575\\
&160.5135369&19 4 15 19 - 18 4 14 19&9.200\\
\hline
&168.9718077&20 4 16 20 19 4 15 20&9.250\\
&168.878091& 20 5 16 20 - 19 5 15 19&9.00\\
&177.3296583& 21 5 16 20 - 20 5 16 19&9.350\\
  163-211 (ALMA Band 5) &185.7821453&22 5 18 23 - 21 5 17 22&9.550\\
&185.8934615&22 4 18 22-21 4 17 22&8.350\\
&194.23659&23 5 19 22 - 22 5 17 22 &10.520\\
&202.6910348& 24 5 20 23 - 23 5 19 24&9.770\\
\enddata
\tablecomments{
$^a$ For the transitions with the same J$^\prime$ k$_a^\prime$ k$_c^\prime$ - $J ^{\prime\prime}$ k$_a^{\prime\prime}$ k$_c^{\prime\prime}$ 
but having different vibrational state, please see the catalog in the \url{propanimine.tar.gz} package included with this article.}
\end{deluxetable*}

\begin{deluxetable*}{cccc}
\tablecaption{Non-LTE Modeling Line Parameters of Trans-ethylamine.}
\tablewidth{0pt}
\tabletypesize{\scriptsize} 
\tablehead{
\colhead{\bf Frequency range (GHz)} & \colhead{\bf Frequency (GHz)$^a$} & \colhead{\bf Transition (J$^\prime$ k$_a^\prime$ k$_c^\prime$ v$^\prime$ - J$^{\prime\prime}$ k$_a^{\prime\prime}$ k$_c^{\prime\prime}$ v$^{\prime\prime}$)} & \colhead{\bf Intensity (K)}
}
\startdata
 & 31.4005744 & 5 1 4 0 - 5 0 5 0 & 0.0108\\
    & 33.06739330 & 2 0 2 0 - 1 0 1 0&0.0096\\
  31-45 (ALMA Band 1)    & 34.98500270 & 6 1 5 0 - 6 0 6 0 & 0.0129 \\
      & 39.42887460 & 7 1 6 1 - 7 0 7 1   &0.0152 \\
   & 44.80634280 & 8 1 7 0 - 8 0 8 0 &0.0173 \\
   \hline
& 80.24199500 & 5 1 5 0 - 4 1 4 0& 0.0534\\
   &82.16878400 & 5 0 5 0 - 4 0 4 0& 0.0569\\
   67-90 (ALMA Band 2)    & 82.67430100 & 5 2 4 0 - 4 2 3 0 & 0.0483\\
and   & 83.2428700 & 5 2 3 1 - 4 2 2 1 &0.0486 \\
84-116 (ALMA Band 3)      & 84.98078500 & 5 1 4 1 - 4 1 3 1 &0.0566\\
   & 98.30233700& 6 0 6 0 - 5 0 5 0& 0.0796\\
  &101.886754 &6 1 5 0 - 5 1 4 0&0.0806 \\
 & 112.1583690 &7 1 7 0 - 6 1 6 0&0.1016 \\
  &114.294235 & 7 0 7 0 - 6 0 6 0  & 0.1052\\
  & 115.604884 & 7 2 6 0 - 6 2 5 0 & 0.0996\\
   \hline
   & 143.9269253 & 9 1 9 0 - 8 1 8 0& 0.1586\\
   & 145.871741 & 9 0 9 0 - 8 0 8 0 & 0.1620\\
 125-163 (ALMA Band 4)   &148.397949 &9 2 8 1 - 8 2 7 1&0.1598\\
    &159.753196 & 10 1 10 0 - 9 1 9 0& 0.1878\\
  & 161.49643100 & 10 0 10 0 - 9 0 9 0 & 0.1914 \\
   \hline
  & 192.561480 & 12 0 12 0 - 11 0 11 0& 0.2432\\
  & 197.25631220 & 12 2 11 1 - 11 2 10 1 & 0.2436\\
163-211 (ALMA Band 5)  & 201.671390 &12 1 11 0 - 11 1 10 0&0.2587\\
  &203.166657 &12 2 10 1 - 11 2 9 1 & 0.2534 \\
  & 208.05354700 &13 0 13 0 - 12 0 12 0& 0.2565\\
\enddata
\tablecomments{
$^a$ For the transitions with same J$^\prime$ k$_a^\prime$ k$_c^\prime$ - $J ^{\prime\prime}$ k$_a^{\prime\prime}$ k$_c^{\prime\prime}$  but having different
vibrational state please see the cat file available in \url{https://www.astro.uni-koeln.de/cdms/catalog}.}
 \end{deluxetable*} 
 
 \begin{deluxetable*}{cccc}
\tablecaption{Non-LTE Modeling Line Parameters of (1Z)-1-Propanimine.}
\tablewidth{0pt}
\tabletypesize{\scriptsize}
\tablehead{
\colhead{\bf Frequency range (GHz)} & \colhead{\bf Frequency (GHz)$^a$} & \colhead{\bf Transition (J$^\prime$ k$_a^\prime$ k$_c^\prime$ v$^\prime$ - J$^{\prime\prime}$ k$_a^{\prime\prime}$ k$_c^{\prime\prime}$ v$^{\prime\prime}$)} & \colhead{\bf Intensity (K)}
}
\startdata
 & 33.51992460 & 4 1 4 5 - 3 1 3 4 & 0.1192\\
  & 33.78887240 & 4 0 4 5 - 3 0 3 4&0.1269\\
31-45 (ALMA Band 1)    & 41.89797460 & 5 1 5 6 - 4 1 4 5& 0.1925\\
 & 42.25035000 & 5 2 3 6 - 4 2 2 5 &0.1752\\
 & 42.58124550 & 5 1 4 6 - 4 1 3 5   &0.1978 \\
   \hline
  & 76.62578210 & 9 1 8 10 - 8 1 7 9& 0.6728\\
   &83.76466100 & 10 1 10 11 - 9 1 9 10 & 0.7793 \\
  67-90 (ALMA Band 2) & 84.36429660 & 10 0 10 11 - 9 0 9 10  & 0.7906\\
and   & 84.56506570 & 10 2 8 11 - 9 2 7 10&0.7710\\
84-116 (ALMA Band 3)    & 85.13188840 & 10 1 9 1 - 9 1 8 10   &0.8215\\
   & 93.63540110 & 11 1 10 12 - 10 1 9 11& 0.9333\\
    &100.49611750 &12 1 12 13 - 11 1 11 12&0.9232\\
      & 101.17028780 &12 0 12 13 - 11 0 11 12&0.9826\\
     &102.13600470 & 12 1 11 13 - 11 1 10 12& 0.9689\\
    & 109.56052950 & 13 0 13 14 - 12 0 12 13& 0.9616\\
   \hline
   & 125.51588020 & 8 2 7 9 - 7 1 6 8& 0.0591\\
   & 126.31319130 & 15 0 15 16 - 14 0 14 15 & 0.6378\\
   125-163 (ALMA Band 4)  &127.00092650 & 15 2 13 16 - 14 2 12 15&0.3568\\ 
  &127.61694790 & 15 1 14 15 - 14 1 13 14 & 0.3922 \\
  & 134.67500350 & 16 0 16 17 - 15 0 15 16   & 0.3738 \\
   \hline
    & 175.64941420 & 9 3 7 10 - 8 2 6 9& 0.09671 \\
   & 175.80109580 & 9 3 6 10 - 8 2 7 9 & 0.09373\\
   163-211 (ALMA Band 5) & 184.02040330 &10 3 8 11 - 9 2 7 10   &0.09697\\
   & 184.25941520 &10 3 7 10 - 9 2 8 9& 0.09647\\
    & 207.19739490 &8 4 5 9 - 7 3 4 8   & 0.09612\\
\enddata
\tablecomments{
$^a$ For the transitions with same J$^\prime$ k$_a^\prime$ k$_c^\prime$ - $J ^{\prime\prime}$ k$_a^{\prime\prime}$ k$_c^{\prime\prime}$ but having different
vibrational state, please see the catalog in the \url{propanimine.tar.gz} package included with this article.}
\end{deluxetable*}


\begin{thebibliography}{99}
\bibitem[\protect\citeauthoryear{Altwegg et al.}{2016}]{altw16}
Altwegg, K., Balsiger. H., Bar-Nun. A., et al. 2016, SciA, 2, e1600285
\bibitem[\protect\citeauthoryear{Bak \& Skaarup}{1971}]{bak71}
Bak. B., \& Skaarup. S. 1971, JMoSt, 10, 385
\bibitem[\protect\citeauthoryear{Brown et al.}{1980}]{brow80}
Brown, R. D., Godfrey, P. D., \& Winkler, D. A. 1980, AJCh, 33, 1
\bibitem[\protect\citeauthoryear{Chakrabarti \& Chakrabarti}{2000a}]{chak00a}
Chakrabarti, S., \& Chakrabarti, S. K. 2000a, A\&A, 354, L6
\bibitem[\protect\citeauthoryear{Chakrabarti \& Chakrabarti}{2000b}]{chak00b}
Chakrabarti, S. K., \& Chakrabarti, S. 2000b, InJPh, 74B, 97
\bibitem[\protect\citeauthoryear{Chakrabarti et al.}{2006a}]{chak06a}
Chakrabarti, S. K., Das, A., Acharyya, K., \& Chakrabarti, S. 2006a, A\&A, 457, 167
\bibitem[\protect\citeauthoryear{Chakrabarti et al.}{2006b}]{chak06b}
Chakrabarti, S. K., Das, A., Acharyya, K., \& Chakrabarti, S. 2006b, BASI, 34, 299
\bibitem[\protect\citeauthoryear{Chakrabarti et al.}{2015}]{chak15}
Chakrabarti, S. K., Majumdar, L., Das, A., \& Chakrabarti, S. 2015, Ap\&SS, 357, 90
\bibitem[\protect\citeauthoryear{Charnley et al.}{1995}]{char95}
Charnley, S. B., Kress, M. E., Tielens, A. G. G. M., \& Millar, T. J. 1995, ApJ, 448, 232

\bibitem[\protect\citeauthoryear{Cronin \& Chang}{1993}]{cron93}
Cronin, J. R., \& Chang, S. 1993, The Chemistry of Life’s Origins, ed. J.
Greenberg, M., Mendoza-Gomez, C. X., Pirronello, V. (Dordrecht: Kluwer), 416, 209

\bibitem[\protect\citeauthoryear{Curtiss et al.}{1997}]{curt97}
Curtiss, L. A., Raghavachari, K., Redfern, P. C., \& Pople, J. A. 1997, JChPh, 106, 1063  
\bibitem[\protect\citeauthoryear{Curtiss et al.}{2007}]{curt07}
Curtiss, L. A., Redfern, P. C., \& Raghavachari, K. 2007a, JChPh, 126, 084108
\bibitem[\protect\citeauthoryear{Das et al.}{2008a}]{das08a}
Das, A., Chakrabarti, S. K., Acharyya K., \& Chakrabarti, S. 2008a, NewA, 13, 457
\bibitem[\protect\citeauthoryear{Das et al.}{2008b}]{das08b}
Das, A., Acharyya, K., Chakrabarti, S. \& Chakrabarti, S. K., 2008b, A\&A, 486, 209
\bibitem[\protect\citeauthoryear{Das et al.}{2010}]{das10}
Das, A., Acharyya, K. \& Chakrabarti, S. K., 2010, MNRAS 409, 789
\bibitem[\protect\citeauthoryear{Das \& Chakrabarti}{2011}]{das11}
Das, A., \& Chakrabarti, S. K. 2011, MNRAS, 418, 545
\bibitem[\protect\citeauthoryear{Das et al.}{2013a}]{das13a}
Das, A., Majumdar, L., Chakrabarti, S. K., \& Chakrabarti, S. 2013a, NewA, 23, 118
\bibitem[\protect\citeauthoryear{Das et al.}{2013b}]{das13b}
Das, A., Majumdar, L., Chakrabarti, S. K., Saha, R., \& Chakrabarti, S. 2013b, MNRAS, 433, 3152
\bibitem[\protect\citeauthoryear{Das et al.}{2015a}]{das15a}
Das, A., Majumdar, L., Chakrabarti, S. K., \& Sahu, D. 2015a, NewA, 35, 53
\bibitem[\protect\citeauthoryear{Das et al.}{2015b}]{das15b}
Das, A., Majumdar, L., Sahu, D., Gorai, P., Sivaraman, B., \& Chakrabarti, S. K. 2015b, ApJ, 808, 21
\bibitem[\protect\citeauthoryear{Das et al.}{2016}]{das16}
Das, A., Sahu, D., Majumdar, L., \& Chakrabarti, S. K. 2016, MNRAS, 455, 540
\bibitem[\protect\citeauthoryear{Elsila et al.}{2007}]{elsi07}
Elsila, J. E., Dworkin, J. P., Bernstein, M. P., Martin, M.P., \& Sandford, S. A. 2007, ApJ, 660, 911
\bibitem[\protect\citeauthoryear{Etim et al.}{2016}]{etim16}
Etim, E., Gorai, P., Das, A., Chakrabarti, S. K. \& Arunan, E. 2016, ApJ, 832, 144
\bibitem[\protect\citeauthoryear{Etim et al.}{2017}]{etim17}
Etim, E., Gorai, P., Das, A. \& Arunan, E. 2017, EPJD, 71, 86
\bibitem[\protect\citeauthoryear{Eyring}{1935}]{eyri35}
Eyring, H. 1935, JChPh, 3, 107
\bibitem[\protect\citeauthoryear{Fischer \& Botshor}{1982}]{fisc82}
Fischer, E., \& Botskor, I. 1982, JMoSp, 91, 116
\bibitem[\protect\citeauthoryear{Fischer \& Botshor}{1984}]{fisc84}
Fischer, E., \& Botskor, I. 1984, JMoSp, 104, 226 
\bibitem[\protect\citeauthoryear{Fortman et al.}{2014}]{fort14}
Fortman, S. M., Neese, C. F., \& De Lucia, F. C. 2014, ApJ, 782, 75
\bibitem[\protect\citeauthoryear{Fourikis et al.}{1974}]{four74}
Fourikis, N., Takagi, K., \& Morimoto, M. 1974, ApJL, 191, L139
\bibitem[\protect\citeauthoryear{Frenkel et al.}{1994}]{fren94}
Frenkel, M., Marsh, K. N., Wilhoit, R. C., et al. 1994, 
Thermodynamics of Organic Compounds in the Gas State, Thermodynamics Research Center, Vol. 1, CRC Press.
\bibitem[\protect\citeauthoryear{Frisch et al.}{2013}]{fris13}
Frisch, M. J., Trucks, G. W., Schlegel, H. B., et al. 2013, Gaussian 09, Revision D.01, Gaussian, Inc., Wallingford CT.
\bibitem[\protect\citeauthoryear{Garrod \& Herbst}{2006}]{garr06}
Garrod, R. T., \& Herbst, E. 2006, A\&A, 457, 927
\bibitem[\protect\citeauthoryear{Garrod, Wakelam \& Herbst}{2007}]{garr07}
Garrod, R. T., Wakelam, V., \& Herbst, E. 2007, A\&A, 467, 1103
\bibitem[\protect\citeauthoryear{Garrod}{2013}]{garr13}
Garrod, R. T. 2013, ApJ, 765, 60

\bibitem[\protect\citeauthoryear{Glavin et al.}{2008}]{glav08}
Glavin, D. P., Dworkin, J. P., \& Sandford, S. A. 2008, M\&PS, 43(1$\textendash$2), 399
\bibitem[\protect\citeauthoryear{Godfrey et al.}{1973}]{godf73}
Godfrey, P. D., Brown, R. D., Robinson, B. J., \& Sinclair, M. W. 1973, ApJ, 13, L119
\bibitem[\protect\citeauthoryear{Gorai et al.}{2017a}]{gora17a}
Gorai, P., Das, A., Das, A., Sivaraman, B., Etim, E. E., \& Chakrabarti, S. K. 2017a, ApJ, 836, 70
\bibitem[\protect\citeauthoryear{Gorai et al.}{2017b}]{gora17b}
Gorai, P., Das, A., Majumdar, L., Chakrabarti, S. K., Sivaraman, B., \& Herbst, E. 2017b, MolAs, 6, 36
\bibitem[\protect\citeauthoryear{Graninger et al.}{2014}]{gran14}
Graninger, D.M., Herbst, E., \"{O}berg, K.I., \& Vasyunin, A.I. 2014, ApJ, 787(1), 74
\bibitem[\protect\citeauthoryear{Hamada et al.}{1986}]{hama86}
Hamada, Y., Tsuboi, M., Yamanouchi, K., \& Kuchitsu, K. 1986, JMoSt, 146, 253
\bibitem[\protect\citeauthoryear{Hasegawa, Herbst \& Leung}{1992}]{hase92}
Hasegawa, T., Herbst, E., \& Leung, C.M. 1992, ApJ, 82, 167
\bibitem[\protect\citeauthoryear{Hendricksen \& Harmony}{1969}]{hend69}
Hendricksen, D., \& Harmony, M. 1969, JChPh, 51, 700

\bibitem[\protect\citeauthoryear{Herbst}{2006}]{herb06}
Herbst, E. 2006, in Springer Handbook of Atomic, Mol. \& Opt. Phys., ed. G.W.F. Drake (New York: Springer), 561

\bibitem[\protect\citeauthoryear{Herzberg}{1966}]{herz66}
Herzberg, G., Molecular spectra and molecular structure. 
Vol. 3: Electronic spectra and electronic structure of polyatomic molecules, New York: Van Nostrand, Reinhold, 1966.

\bibitem[\protect\citeauthoryear{Holtom et al.}{2005}]{holt05}
Holtom, P. D., Bennett, C. J., Osamura, Y., Mason, N. J., \& Kaiser, R. I. 2005, ApJ, 626, 940


\bibitem[\protect\citeauthoryear{Johnson \& Lovas.}{1972}]{john72}
Johnson, D. R., \& Lovas, F. J. 1972, CPL, 15, 65
\bibitem[\protect\citeauthoryear{Kaifu et al.}{1974}]{kaif74}
Kaifu, N., Morimoto, M., Nagane, K., Akabane, K., Iguchi, T., \& Takagi, K. 1974, ApJL, 191, L135

\bibitem[\protect\citeauthoryear{Lakard}{2003}]{laka03}
Lakard, B. 2003, Int. Elec. Conf. Mol. Des., Bochem press, http://www.biochempress.com.

\bibitem[\protect\citeauthoryear{Landolt et al.}{1974}]{land74}
Landolt, H., B$\ddot{o}$rnstein, R., \& Hellwege, K. H. 1974, 
Molecular Constants from Microwave, Molecular Beam, and Electron Spin Resonance Spectroscopy, Vol. 6, Springer.

\bibitem[\protect\citeauthoryear{Leung, Herbst \& Huebner}{1984}]{leun84}
Leung, C. M., Herbst, E., \& Huebner, W. F. 1984, ApJS, 56, 231

\bibitem[\protect\citeauthoryear{Lias et al.}{2005}]{lias05}
Lias, S., Linstrom, P. J., Mallard, W. G. (Eds.), NIST Chemistry WebBook, NIST Standard Reference 
Database Number 69, http://webbook.nist.gov, Gaithersburg MD, 2005.

\bibitem[\protect\citeauthoryear{Loomis et al.}{2013}]{loom13}
Loomis, R. A., Zaleski, D. P., Steber, A. L., et al. 2013, ApJL, 765, L9
\bibitem[\protect\citeauthoryear{Lovas et al.}{1980}]{lova80}
Lovas, F. J., Suenram, R. D., Johnson, D. R., Clark, F. O., \& Tiemann, E. 1980, JChPh, 72, 4964
\bibitem[\protect\citeauthoryear{Majumdar et al.}{2012}]{maju12}
Majumdar, L., Das, A., Chakrabarti, S. K., \& Chakrabarti, S. 2012, RAA, 12, 1613
\bibitem[\protect\citeauthoryear{Majumdar et al.}{2013}]{maju13}
Majumdar, L., Das, A., Chakrabarti, S. K., \& Chakrabarti, S. 2013, NewA, 20, 15
\bibitem[\protect\citeauthoryear{Majumdar et al.}{2014a}]{maju14a}
Majumdar, L., Das, A., \& Chakrabarti, S. K. 2014a, A\&A, 562, A56
\bibitem[\protect\citeauthoryear{Majumdar et al.}{2014b}]{maju14b}
Majumdar, L., Das, A., \& Chakrabarti, S. K. 2014b, ApJ, 782, 73

\bibitem[\protect\citeauthoryear{Margul$\grave{e}$s et al.}{2015}]{marg15}
Margul$\grave{e}$s, L., Motiyenko, R. A., Guillemin, J. C., \& Cernicharo, J., 70$^{th}$ International 
Symposium on Molecular Spectroscopy: June 22-26, 2015 at The University of Illinois at Urbana-Champaign. Talk RI07,
International Symposium on Molecular Spectroscopy (CHAMPAIGN-URBANA, ILLINOIS)

\bibitem[\protect\citeauthoryear{McElroy et al.}{2013}]{mcel13}
McElroy, D., Walsh, C., Marckwick, A. J., Cordiner, M. A., Smith, K., Millar, T. J. 2013, A\&A, 550. 36
\bibitem[\protect\citeauthoryear{McMillan et al.}{2014}]{mcml14}
McMillan, J. P., Fortman, S. M., Neese, C. F., \& De Lucia, F. C. 2014, ApJ, 795, 56
\bibitem[\protect\citeauthoryear{Mehrotra et al.}{1977}]{mehr77}
Mehrotra, S. C., Griffin, L. L., Britt, C. O., \& Boggs. J.E. 1977, JMoSp, 64, 244
\bibitem[\protect\citeauthoryear{M\"{u}ller et al.}{2016}]{mull16}
M\"{u}ller, H. S. P., Belloche, A., Xu, L., et al. 2016, A\&A, 587, A92
\bibitem[\protect\citeauthoryear{Nelson et al.}{1967}]{nels67}
Nelson Jr, R. D., Lide Jr, D. R., \& Maryott, A. A. 1967,
Selected Values of electric dipole moments for molecules in the gas phase, (No. NSRDS-NBS10). 
National Standard Reference Data System.

\bibitem[\protect\citeauthoryear{Ohishi et al.}{2017}]{ohis17}
Ohishi, M., Suzuki, T., Hirota, T., Saito, M. \& Kaifu, N. 2017, arXiv:1708.06871

\bibitem[\protect\citeauthoryear{Osmont et al.}{2007}]{osmo07}
Osmont, A., Catoire, L., G$\ddot{o}$kalp, I., \& Yang, V. 2007, CoFl, 151, 262
\bibitem[\protect\citeauthoryear{Pearson et al.}{1977}]{pear77}
Pearson Jr, R., \& Lovas, F.J. 1977, JChPh, 66(9), 4149
\bibitem[\protect\citeauthoryear{Pickett}{1991}]{pick91}
Pickett, H. M. 1991, JMoSp, 148, 371
\bibitem[\protect\citeauthoryear{Quan et al.}{2010}]{quan10}
Quan, D., Herbst, E., Osamura, Y., \& Roueff, E. 2010, ApJ, 725, 2101
\bibitem[\protect\citeauthoryear{Quan et al.}{2016}]{quan16}
Quan, D., Herbst. E., Corby, J. F., Durr. A., \& Hassel, G. 2016, ApJ, 824, 129
\bibitem[\protect\citeauthoryear{Ruaud et al.}{2016}]{ruau16}
Ruaud, M., Wakelam, V., Hersant, F. 2016, MNRAS, 459, 3756
\bibitem[\protect\citeauthoryear{Sahu et al.}{2015}]{sahu15}
Sahu, D., Das, A., Majumdar, L., \& Chakrabarti, S. K. 2015, NewA, 38, 23
\bibitem[\protect\citeauthoryear{Sastry \& Curl}{1964}]{sast64}
Sastry, K. V. L. N., \& Curl Jr, R. F. 1964, JChPh, 41(1), 77

\bibitem[\protect\citeauthoryear{Sil et al.}{2017}]{sil17}
Sil, M., Gorai, P., Das, A., Sahu, D., \& Chakrabarti, S. K. 2017, EPJD, 71(2), 45
\bibitem[\protect\citeauthoryear{Sharma \& Chandra}{2001}]{shar01}
Sharma, A. K., \& Chandra, S. 2001, A\&A, 376, 333

\bibitem[\protect\citeauthoryear{Suzuki et al.}{2016}]{suzu16}
Suzuki, T., Ohishi. M., Hirota. T., Saito, M., Majumdar. L, \& Wakelam, V. 2016, ApJ, 825, 1
\bibitem[\protect\citeauthoryear{Takagi \& Kojima}{1973}]{taka73}
Takagi, K., \& Kojima, T. 1973, ApJL, 181, L91
\bibitem[\protect\citeauthoryear{Theule et al.}{2011}]{theu11}
Theule, P., Borget, F., Mispelaer, F., et al. 2011, A\&A, 534, A64
\bibitem[\protect\citeauthoryear{Van der Tak et al.}{2007}]{vand07}
Van der Tak, F. F. S., Black, J. H., Sch$\ddot{o}$ier, F. L., Jansen, D. J., van Dishoeck, E. F. 2007, A\&A 468, 627
\bibitem[\protect\citeauthoryear{Vasyunin \& Herbst}{2013}]{vasy13}
Vasyunin, A. I., \& Herbst, E. 2013, ApJ, 769, 34
\bibitem[\protect\citeauthoryear{Vuitton et al.}{2006}]{vuit06}
Vuitton, V., Yelle, R. V., Anicich, V. G. 2006, ApJ, 647, 175
\bibitem[\protect\citeauthoryear{Wollrab \& Laurie}{1968}]{woll68}
Wollrab, J. E., \& Laurie, V. W. 1968, JChPh, 48(1), 5058
\bibitem[\protect\citeauthoryear{Woon}{2002}]{woon02}
Woon, D. E. 2002, ApJL, 571, L177
\end{thebibliography}
\end{document}